\documentclass[a4paper,10pt]{article}
\pdfoutput=1 

\usepackage{jheppub} 

\usepackage[T1]{fontenc} 

\usepackage{amsfonts}
\usepackage{amsmath}
\usepackage{amsthm}
\usepackage{amssymb}
\usepackage{array}
\usepackage{dcolumn}
\usepackage{placeins}
\usepackage[svgnames]{xcolor}

\usepackage{graphics}
\usepackage{tikz}
\usepackage{graphicx}
\usetikzlibrary{positioning}
\usepackage{caption}
\usepackage{subcaption}
\usepackage{adjustbox}
\usepackage{gensymb}
\usepackage{breqn}
\usepackage{braket}
\usepackage{enumitem}
\usepackage{makecell}
\usepackage{ragged2e}
\usepackage{comment}
\usepackage{url}
\usepackage{placeins}
\usepackage{blindtext}
\usepackage{float}
\usepackage{hyperref}

\preprint{MI-HET-775, KIAS-P22043}

\newcommand\be{\begin{equation}}
\newcommand\ee{\end{equation}}
\newcommand\bea{\begin{eqnarray}}
\newcommand\eea{\end{eqnarray}}

\title{\boldmath Non-standard neutrino interactions in light mediator models at reactor experiments}

\author[a]{Bhaskar Dutta,}
\author[a, b]{Sumit Ghosh,}
\author[c, d]{Tianjun Li,}
\author[a]{Adrian Thompson,}
\author[a]{and Ankur Verma}

\affiliation[a]{Mitchell Institute for Fundamental Physics and Astronomy, Department of Physics  and Astronomy, Texas A$\&$M University, College Station, Texas 77843,  USA}

\affiliation[b]{School of Physics, Korea Institute for Advanced Study, Seoul 02455, Korea}

\affiliation[c]{CAS Key Laboratory of Theoretical Physics, Institute of Theoretical Physics,
Chinese Academy of Sciences, Beijing, 100190, People’s Republic China}

\affiliation[d]{School of Physical Sciences, University of Chinese Academy of Sciences,
Beijing 100049, People’s Republic China}

\emailAdd{dutta@physics.tamu.edu}
\emailAdd{ghosh@kias.re.kr}
\emailAdd{tli@itp.ac.cn}
\emailAdd{thompson@tamu.edu}
\emailAdd{averma1@tamu.edu}

\abstract{ Compared to other neutrino sources, the huge anti-neutrino fluxes at nuclear reactor based experiments empower us to derive stronger bounds on non-standard interactions of neutrinos with electrons mediated by light scalar/vector mediators. At neutrino energy around $200$~keV reactor anti-neutrino flux is at least an order of magnitude larger compared to the solar flux. The atomic and crystal form factors of the detector materials  related to the details of the atomic structure becomes relevant at this energy scale as the momentum transfers would be small. Non-standard neutrino-electron interaction mediated by light scalar/vector mediator arises naturally in many low-scale models. We also propose one such new model with a light scalar mediator. Here, we investigate the parameter space of such low-scale models in reactor based neutrino experiments with low threshold Ge and Si detectors, and find the prospect of probing/ruling out the relevant parameter space by finding the projected sensitivity at $90 \%$ confidence level  by performing a $\chi^2$-analysis. We find that  a detector capable of discriminating between electron recoil and nuclear recoil signal down to a very low threshold such as $5$~eV placed in reactor based experiment would be able to probe a larger region in parameter space compared to the previously explored region. A  Ge (Si) detector with $10$~kg-yr exposure and 1 MW reactor anti-neutrino flux would be able to probe the scalar and vector mediators with masses below 1 keV for coupling products $\sqrt{g_\nu g_e}$ $\sim$ $1 \times 10^{-6}~(9.5 \times 10^{-7}) ~{\rm and}~ 1\times 10^{-7} ~(8\times 10^{-8})$, respectively.

}

\keywords{NSI, Reactor Neutrinos}

\begin{document} 
\maketitle
\flushbottom

\section{Introduction}
\label{sec:intro}

The solar, atmospheric, reactor, and accelerator neutrino experiments have established that three neutrinos oscillate and at least two neutrinos in the Standard Model (SM) are massive~\cite{Super-Kamiokande:1998kpq, SNO:2002tuh, KamLAND:2002uet, T2K:2011ypd, DoubleChooz:2011ymz, DayaBay:2012fng, RENO:2012mkc}. However, the origin of neutrino masses and the variety of neutrino interactions still remain to be explained. Many models have been proposed  to explain the tiny non-zero masses of neutrinos with the scales of new physics covering a wide range of values~\cite{Minkowski:1977sc, Yanagida:1979as, Gell-Mann:1979vob, Mohapatra:1979ia, Mohapatra:1980yp, Cheng:1980qt, Foot:1988aq, Schechter:1980gr, Chikashige:1980ui, Ballesteros:2016euj, Barbieri:1979hc, Grossman:1999ra, Huber:2003sf, Agashe:2015izu, Mohapatra:1986aw, Mohapatra:1986bd, Kersten:2007vk, Zee:1980ai, Wolfenstein:1980sy, Zee:1985rj, Babu:1988ki, Ma:1998dn, Ma:2006km, Ma:2007yx, Dev:2012sg, Cao:2017xgk, Dutta:2018qei, Ibarra:2018dib, Cacciapaglia:2020psm}. The well-established phenomena of neutrino oscillations and non-zero masses establish the fact that there exists new physics in the neutrino sector beyond the SM. The non-standard interactions in the neutrino sector are a well- motivated phenomenological approach to understand the new physics in the neutrino sector~\cite{Wolfenstein:1977ue, Wolfenstein:1979ni, Bergmann:1997mr, Valle:1987gv, Roulet:1991sm, Krastev:1997cp, Miranda:2015dra}.  So far, there is no direct experimental observation that shows new interactions of neutrinos beyond the SM weak interactions.  If they exist, they would show some signals in the ongoing and upcoming neutrino experiments, and thus would directly indicate the presence of new physics beyond the SM.

Recently, the low-scale neutrino models containing non-standard interaction mediated by light mediators have been receiving a lot of attention. Such models can naturally accommodate neutrino-electron interactions beyond the SM weak interactions. There exist many vector models in the literature~\cite{Foot:1990mn, He:1990pn, He:1991qd, Dutta:2019fxn, Dutta:2022qvn}, but there is a dearth of light scalar mediator models. We propose one light scalar model by extending the SM scalar sector by a doublet, a triplet and a singlet scalar. We will use these models as the base models for our analysis. Many ongoing beam-dump based neutrino experiments, e.g., COHERENT~\cite{Akimov:2017ade, Akimov:2018vzs, Akimov:2018ghi, Akimov:2019xdj, Akimov:2020pdx, COHERENT:2021pvd}, CCM~\cite{CCM1, CCM2, CCM:2021leg}, JSNS$^2$~\cite{Harada:2013yaa, Ajimura:2015yux, Harada:2015tcp, Ajimura:2017fld, Rott:2020duk} etc., and reactor based experiments, e.g., GEMMA~\cite{Beda:2013mta}, Borexino~\cite{BOREXINO:2018ohr}, MINER~\cite{MINER:2016igy}, CONUS~\cite{Buck:2020opf}, CONNIE~\cite{Aguilar-Arevalo:2016khx} etc. are investigating these low mass mediator models~\cite{  Cerdeno:2016sfi,Dutta:2020vop,Strauss:2017cuu,Dent:2016wcr}. Since the neutrinos   produce a background for the dark matter direct detection experiments, these new interactions are being investigated at the dark matter experiments, e.g., XENONnT~\cite{XENON:2017lvq, XENON:2022mpc}, LZ~\cite{Akerib:2019fml}, SENSEI~\cite{SENSEI:2020dpa}, etc. Some of these experiments put stringent bounds on the parameter space of these models. In addition to these, the light mediators of these models can also be constrained from astrophysical and cosmological data~\cite{Babu:2019iml, Harnik:2012ni, Escudero:2019gvw}.  The neutrino interactions with low mass mediators also appear to be resolving the so-called Hubble tension~\cite{Escudero:2019gzq, Bernal:2016gxb, Alcaniz:2019kah, Vagnozzi:2019ezj}.

Among various sources of neutrinos available for neutrino experiments, reactor neutrino flux can be larger compared to others for a particular neutrino energy range for  proper settings of the reactor. For example, at $\sim 200-300$~keV neutrino energy range a 1 MW class reactor with source-to-detector distance of 1 m can give one order of magnitude larger flux compared to solar flux. Therefore, one can expect that the reactor based neutrino experiments can probe a much bigger parameter space compared to the solar neutrino based analysis such as XENONnT and Borexino. This is one of the main motivations of this paper.

In this paper, we investigate the parameter space of  neutrino-electron scattering mediated by light mediators in low scale neutrino models utilizing the large anti-neutrino flux from reactor based experiments. The mediators can be scalar, pseudo-scalar, vector, and axial-vector types, but we mainly focus on scalar and vector cases. Similar analysis can be performed for pseudo-scalar and axial-vector mediators as well. The neutrino-electron scattering processes via light mediators have an interesting feature: the cross section becomes larger at low recoil energies, hence getting an enhancement in the rate. Therefore, the low threshold detectors are a good probe to the light mediators. For our analysis purpose, we consider that the detector used in the experiment would have a very low threshold such as 5 eV, and a very low constant background over the SM weak current mediated neutrino-electron scattering events. We further consider that the detector can distinguish events based on whether it is coming from electronic recoil or nuclear recoil. Such capability of the detector turns out to be very fruitful for the ranges of parameters we are considering.

 For light mediators with mass  $\mathcal{O}(1)$ keV, the energy transferred in the scattering falls in sub-keV range and the related distance scale becomes comparable to the atomic scale. Therefore, a free electron approximation, where we treat the atom as something with $Z$ number of free electrons, is not a good choice for our analysis. For proper treatment, one needs to consider the many body interaction picture inside the target atoms and consider the ionization effects. Moreover, the target materials we are considering, {\it i.e.} silicon and germanium, are semiconductor crystals. Below some energy scale the energy band structure of the crystals becomes relevant. The dynamics of many body ionization and the transition between different energy bands inside the crystal  can be encapsulated in atomic and crystal form factors, respectively. We use a combination of ionization form factor and crystal form factor depending on the energy scale.  The crystal form factor  accounts for low energy $\lesssim$ 60 eV transitions from valence to conduction bands. On the other hand, the atomic form factor is used for the transitions to energies $\gtrsim$ 60 eV in the conduction bands from some initial electronic state in the crystal. We use the \texttt{DarkARC} python tool~\cite{catena_atomic_2020} to get the atomic form factors, and compute the crystal form factors using the Fortran program \texttt{EXCEED-DM}~\cite{griffin_extended_2021}.

The rest of the paper is organized as follows: In section~\ref{sec:neutrino-electron_scattering}, we give a brief review of neutrino-electron scattering mediated by light mediators. The relevant ionization and crystal form factors are discussed in section~\ref{sec:formfactor}. In section~\ref{sec:models}, we propose a new light scalar model that can give rise to non-standard interactions between neutrinos and electrons. Few light vector models have been reviewed as well. We present our results in section~\ref{sec:results}. We conclude in section~\ref{sec:conclusion}.

\section{Neutrino-Electron Scattering} \label{sec:neutrino-electron_scattering}

In this section, we discuss the basic framework of NSI with electrons via light mediators arising from some new physics scenarios. The simplest type of NSI, which leads to neutrino-electron scattering at low energies, is through the exchange of light scalar or vector particles.

In the SM, the low energy neutrinos and anti-neutrinos interact with electrons through the elastic scattering process via the charged current (exchange of $W$ boson) and neutral current ( exchange of $Z$ boson). The neutrino-electron scattering cross section in SM is given by 
\bea \label{nu-ecrosssection:SM} \frac{d \sigma_e^{\mathrm{SM}}}{dT} = \frac{G_F^2 m_e}{2 \pi} \left[ (g_v +g_a)^2 + (g_v -g_a)^2 \left( 1-\frac{T}{E_\nu} \right)^2 - (g_v^2-g_a^2) \frac{m_e T}{E_\nu^2} \right] ~,~\eea where $G_F$ is the Fermi constant; $m_e$ is the electron mass; $T$ is the recoil energy of the outgoing electron; and $E_\nu$ is the incoming neutrino energy. The couplings $g_v$ and $g_a$ depend on the flavor of the neutrinos. This is due to the fact that the electron neutrino scattering with electron receives contributions from both $W$-boson and $Z$-boson exchange diagrams, whereas for muon and tao neutrino scattering with electron only the $Z$-boson exchange diagram is possible. For $\nu_e$ and $\bar{\nu}_e$ we have \be g_{v;e}= \frac{1}{2}+ 2 \sin^2\theta_W;~~~~~ g_{a;e}=+\frac{1}{2} ~,~ \ee whereas for $\nu_{\mu, \tau}$ and $\bar{\nu}_{\mu, \tau}$ we have \be g_{v;\mu, \tau}=-\frac{1}{2}+ 2 \sin^2\theta_W;~~~~~ g_{a;\mu, \tau}=-\frac{1}{2} ~,~  \ee where $\theta_W$ is the weak mixing angle with $\sin^2_{\theta_W} = 0.231$~\cite{ParticleDataGroup:2020ssz}. Note that, for the range of electron recoil energy where, $T \ll E_\nu$, the SM differential cross section is constant.

The neutrino-electron scattering cross section deviates from Eq.~\ref{nu-ecrosssection:SM} in the presence of NSI. The new physics contributions will be different for light scalar and light vector mediators. In the following, we summarize the new physics contributions to the neutrino-electron scattering cross section for light scalar and light vector particle exchange processes. 

\begin{itemize}
    \item Light scalar mediator: We consider a light scalar particle, $\phi$, with mass $m_\phi$ that couples to both electron and neutrino at tree level with coupling constants $g_{e, \phi}$ and $g_{\nu, \phi}$, respectively. The necessary terms of the low-energy Lagrangian below the electroweak (EW) scale can be written as \be \mathcal{L}_{S} \supset g_{e, \phi} \bar{e} e \phi + g_{\nu, \phi} \bar{\nu} \nu \phi~.~\,\ee The $\bar{\nu} \nu \phi$ interaction term can be realized in different ways. One can simply add a right-handed neutrino and a Dirac type interaction term $\bar{\nu}_R \nu_L \phi$. Another possibility is to generate a Majorana type interaction term like $\bar{\nu}^c_L \nu_L \phi$ using a triplet scalar field, which we will show in detail in subsection~\ref{models:scalar}. The new contribution to the differential cross section of  neutrino-electron scattering is given by \begin{equation} \label{nu-ecrosssection:scalar} \frac{d \sigma_{e}}  {d T} -\frac{d \sigma_{e}^{\mathrm{SM}}} { d T }= \frac{{g_{\nu, \phi}}^{2} g_{e, \phi}^{2} T m_{e}^{2}}{4 \pi E_\nu^2 (2 T m_e + m_\phi^2)^2 } ~.~\, \end{equation} Note that, the new contribution to the differential cross section varies as $1/T$ for the range $T > m_\phi$, whereas it decreases proportional to smaller $T$, when $T$ is much smaller than  $m_\phi$. Also note that there is no interference term.
    
    \item Light vector mediator: Here we consider a light vector mediator $A^\prime$ with mass $m_{A^\prime}$. The $A^\prime$ couples to both electron and neutrino at tree level with respective coupling constants  $g_{e, A^\prime}$ and $g_{\nu, A^\prime}$. The low-energy Lagrangian below EW scale is \be \mathcal{L}_{V} \supset g_{e, A^\prime} \bar{e} \gamma^\rho e A^\prime_\rho + g_{\nu, A^\prime} \bar{\nu}_L \gamma^\rho \nu_L A^\prime_\rho ~.~\,\ee The new contribution to the differential cross section can be written as \begin{equation} \frac{d \sigma_{e}}{  d T} -\frac{d \sigma_{e}^{\mathrm{SM}}} {d T} = \frac{\sqrt{2} G_{F} m_{e} g_{v} g_{\nu, A^{\prime}} g_{e, A^{\prime}}}{\pi\left(2 m_{e} T +m_{A^{\prime}}^{2}\right)} +\frac{m_{e} g_{\nu, A^{\prime}}^{2} g_{e,  A^{\prime}}^{2}}{2 \pi\left(2 m_{e} T +m_{A^{\prime}}^{2}\right)^{2}} ~.~\,\end{equation} The new light vector contribution to the differential cross section has an interference term with SM. The cross section first increases by $1/T^2$  when $m_{A^\prime}$ can be neglected compared to $T$, and becomes constant when energy transfer is smaller than the mediator mass.
\end{itemize}

\begin{figure}[tbp]
\centering 
\includegraphics[width=0.9\textwidth]{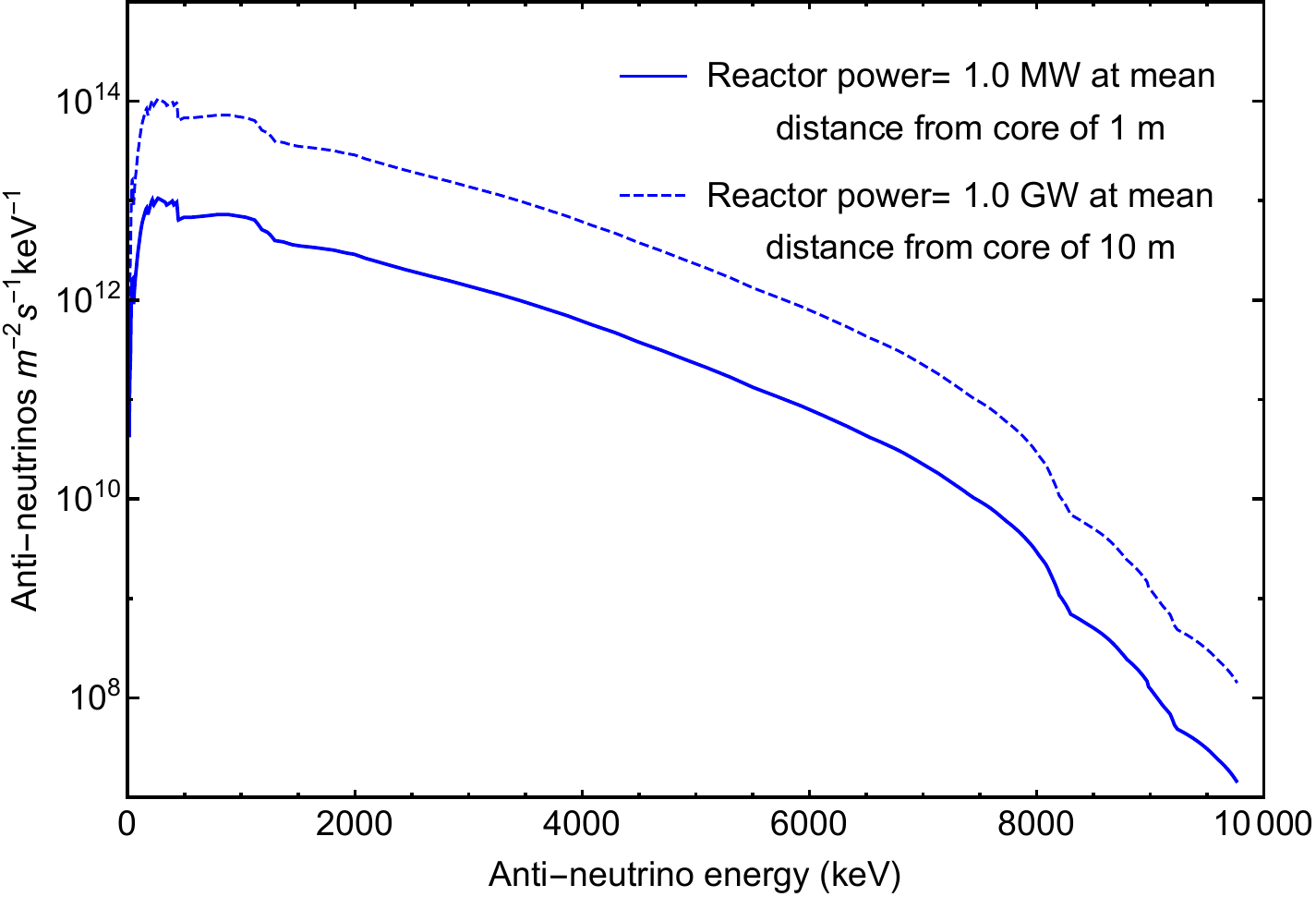}

\captionsetup{justification   = RaggedRight,
             labelfont = bf}
\caption{\label{fig:reactorflux} Anti-neutrino spectrum produced in the nuclear fission process at reactor based neutrino experiment. We use the spectrum corresponding to 1 MW power reactor with core to detector distance 1 meter (solid line). We also show that, by using a 1 GW power reactor with core to detector distance 10 meters, we can get one order of magnitude more flux (Dashed line).}
\end{figure}

For our analysis, we mostly consider the anti-neutrino flux generated from nuclear fission in a  reactor~\cite{Schreckenbach:1985ep, Kopeikin:2012zz}. Specifically, we use the anti-neutrino fission spectrum which is used for the analysis in the MINER experiment at Texas A$\&$M University~\cite{Dutta:2015vwa}. MINER is a reactor based experiment equipped with a low threshold ( $\sim$ 5 eV) cryogenic detector. The reactor is a megawatt (MW) class TRIGA research reactor with movable core stocked with low-enriched ($\sim$ 20 $\%$) ${}^{235} U$. This high resolution detector with a movable core provides an ideal setup to probe neutrino NSI where the recoil energy is very low. The detector can be installed at a baseline separation of 1 m from the reactor core. The thermal power in the reactor is generated by the nuclear fission process. The intrinsic anti-neutrino flux can be obtained by combining the reactor thermal power and the relevant fission fractions. At 1.00 MW reactor power, the extrapolated fission rate is $ 3.1 \times 10^{16} s^{-1}$ with an intrinsic anti-neutrino production rate of $1.9 \times 10^{17} s^{-1}$, and a fission anti-neutrino flux of $\Phi_\nu = 1.5 \times 10^{12} cm^{-2}s^{-1}$ at a mean distance from core of 1 m. Note that, these fluxes are proportional to $1/r^2$, where $r$ is the mean distance between the detector and the reactor core. Therefore the flux is reduced by a factor of 4 (9) for a mean distance of 2 m (3 m). The  anti-neutrino flux used for analysis is shown in  figure~\ref{fig:reactorflux}.

The anti-neutrino flux from MW class reactors has a similar energy profile to the solar neutrino flux~\cite{SNO:2011hxd, Super-Kamiokande:2016yck, BOREXINO:2018ohr} with characteristic neutrino energies $\le 1$ MeV. Both reactor anti-neutrino and solar neutrino flux are dominated by the neutrinos with energy $\mathcal{O}(100)$ keV. At 1.00 MW reactor power  with a core to detector mean distance of 1 m, the anti-neutrino flux gets a peak between 200-300 keV neutrino energy, whereas the solar neutrino flux gets a peak around 200 keV. One advantage of the reactor anti-neutrino flux is that at this keV energy scale it gives one order of magnitude more flux compared to the solar neutrino flux. Experiments like XENONnT and Borexino have already provided constraints on the parameter space of the neutrino NSI using the solar-$pp$ flux. Thus, using the larger anti-neutrino flux from the reactors we expect to see even greater sensitivity of the parameter space. Note that, with a gigawatt (GW) class nuclear reactor we can get 10 times more flux even with a 10 m mean distance between the reactor core and the detector, such as in vIOLETA experiment. Also note that at the typical neutrino energies $\le 200$ keV, the atomic/crystal structure of the target material becomes relevant. Therefore, one needs to incorporate these effects in the rate calculation~\cite{Chen:2014ypv}, which we will discuss in detail in section~\ref{sec:formfactor}.

\begin{figure}[tbp]
\centering 
\includegraphics[width=.45\textwidth]{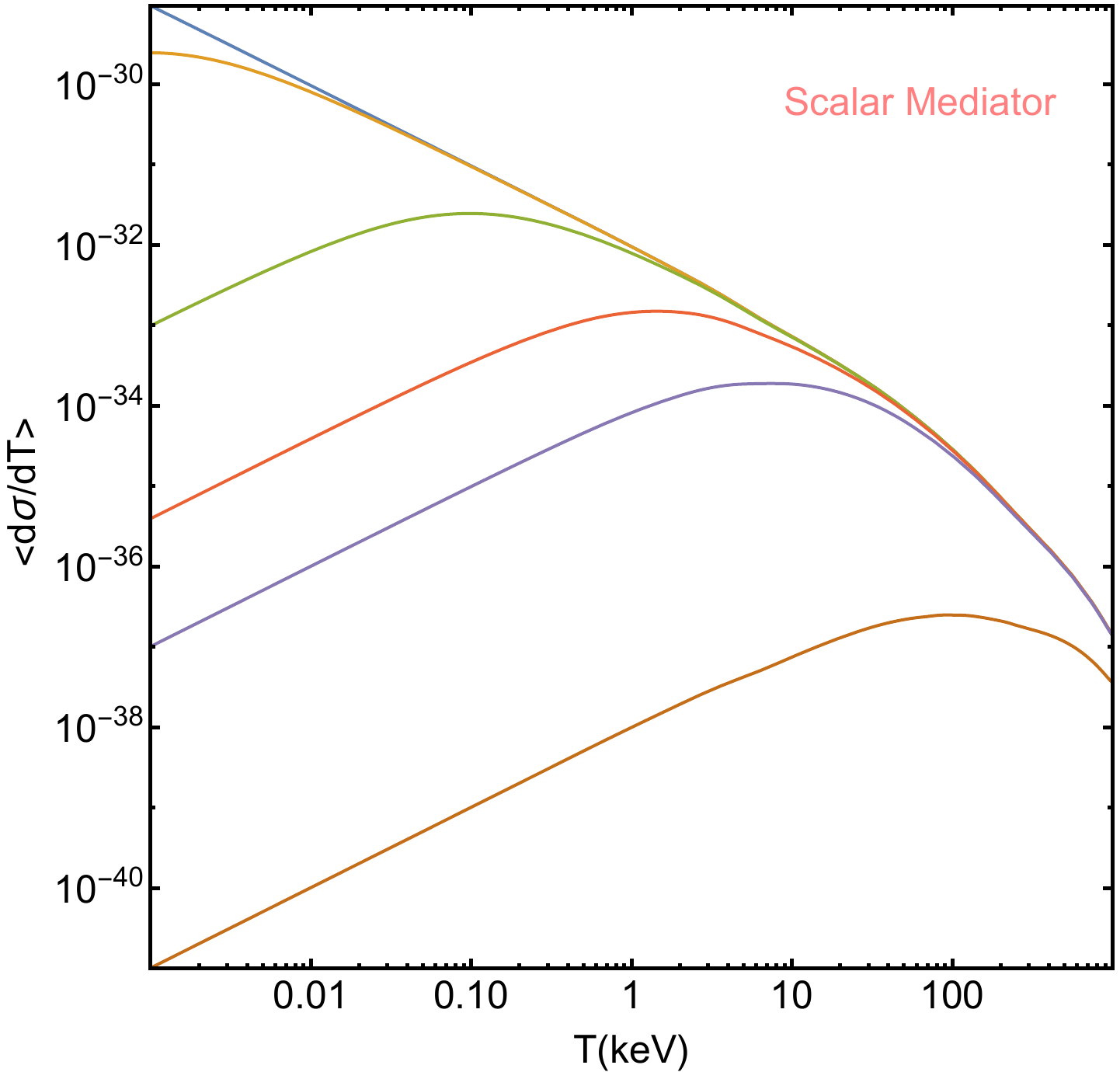}
\hfill
\includegraphics[width=.45\textwidth]{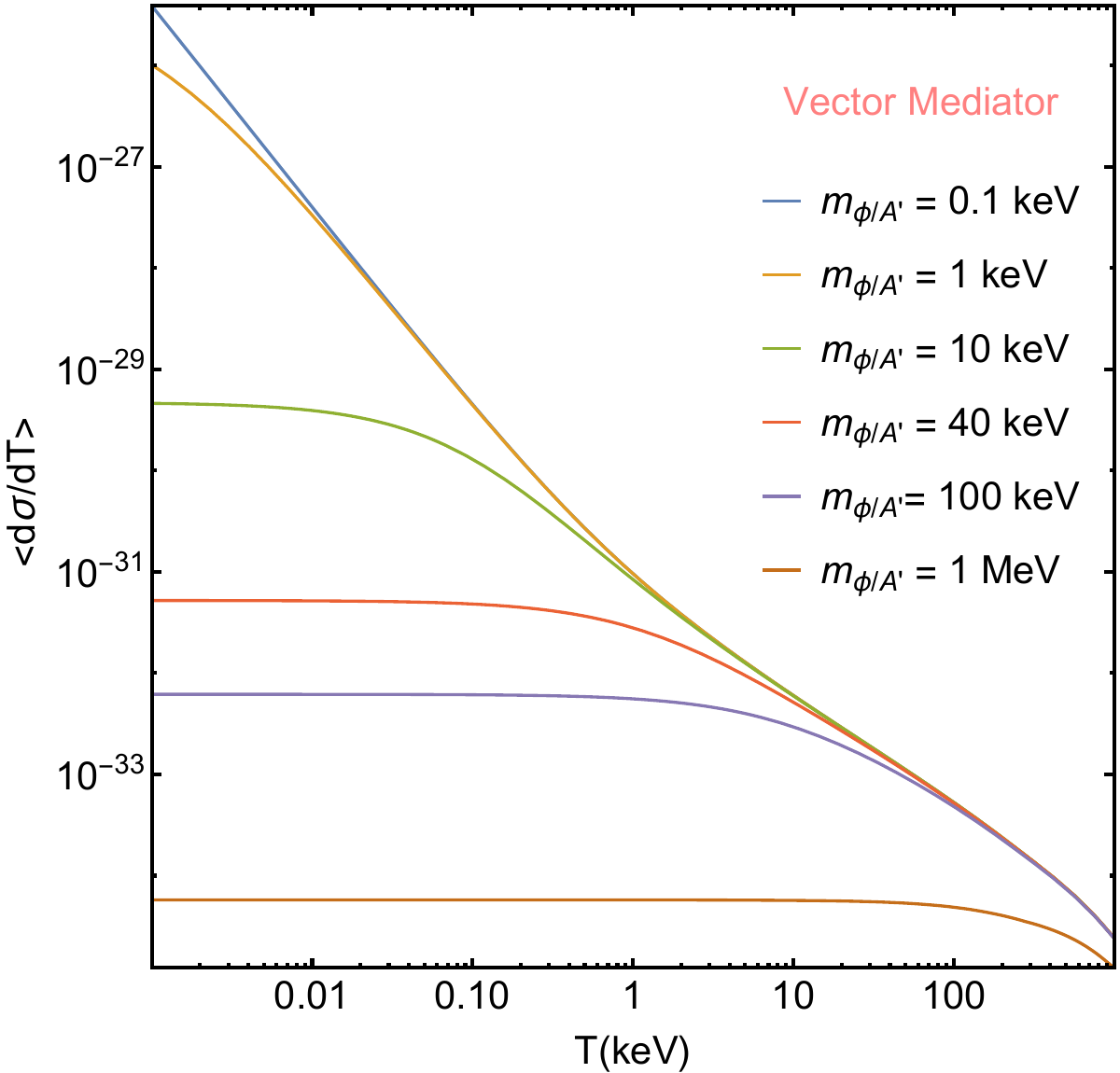}
\captionsetup{justification   = RaggedRight,
             labelfont = bf}
\caption{\label{fig:dsigdT} 
Flux averaged differential cross section. We have used the reactor anti-neutrino flux from figure~\ref{fig:reactorflux}. The left panel is for scalar, and the couplings are: $g_{\nu}=g_{e}=5 \times 10^{-7}$. The vector case is on the right panel, and the benchmark couplings are: $g_{\nu}=g_{e}=1.5 \times 10^{-7}$. }
\end{figure}

In figure~\ref{fig:dsigdT}, we show the flux averaged differential cross-section for different mediator masses of both scalar and vector particles. We have used the reactor anti-neutrino flux from figure.~\ref{fig:reactorflux}. For the scalar case, we choose a benchmark coupling $g_\nu = g_e = 5 \times 10^{-7} $. And for the vector mediator, we take $g_\nu = g_e = 1.5 \times 10^{-7} $.
 The cross sections for the scalar mediated case are first increasing as $1/T$ with decreasing $T$, then attain a peak and further fall down linearly with $T$. On the other hand, the vector mediated cross sections increase proportional to $1/T^2$ when energy transfer is large compared to the mass of the mediator. The cross section then becomes independent of $T$ and go parallel to the x-axis. Note that, for  scalar (vector) mediated cases, the curves attain a peak  (become constant) below $\mathcal{O}(1)$~keV for light mediator of mass $\mathcal{O}(10)$~keV.

\section{Atomic Ionization by Neutrino NSI} \label{sec:formfactor}

The conventional free electron approximation for isolated atoms is expected to be a good approximation when the momentum transferred in the scattering process is large so that the corresponding de Broglie wavelength is much smaller than the typical distance between the electrons in a multi-electron atom. In other words, when the typical scattering energy is high compared to the atomic scales, the atomic effects can be neglected. This case would just correspond to a factor of Z in the scattering rate accounting for the Z electrons in the atom. However, as has been shown in some previous works \cite{Chen:2013iud, Chen:2014ypv, Hsieh:2019hug} that for the incident neutrino energy less than the typical binding momentum of electron $\sim Z m_{e} \alpha$ in a hydrogen like atom, the simple free electron approximation is not accurate. This would correspond to neutrino energies of $\mathcal{O}(100)~{\rm keV}$ which is quite typical of neutrino energies we get from a MW class reactor flux.

For the NSI with light mediators the momentum transferred during an interaction event is small and the differential cross section peaks at very low recoil energies as can be seen in figure~\ref{fig:dsigdT}. As the kinematics in neutrino scattering with sub-keV energy transfer starts to overlap with atomic scales, the atomic binding effects modify the free scattering rate formula. Thus, for getting a better understanding of detector responses at low recoil energies
one must deal with a proper treatment of many electron dynamics in atomic ionization. The effect of this complication can be encoded inside an atomic ionization form factor in the rate calculation. 
Furthermore, for crystal targets like Si/Ge the atoms are not isolated which modifies the outer shell electron wave functions. The outer-shell electronic wavefunctions are delocalized forming a complicated band structure with an energy band gap between the occupied valence bands and unoccupied conduction bands. This modification becomes important at ultra-low energy transfers when most of the transitions will be occurring from the valence band to the conduction band. This is accounted for in the form of a crystal form factor which describes the likelihood that a given momentum transfer will result in a transition from some state in valence band to conduction band.

 Therefore,  the free scattering differential rate formula will be modified after incorporating the form factors. The differential rate can be thought of as a sum of two contributions: rate modified by atomic form factor and crystal form factor.   We briefly describe the atomic and crystal form factors and how  they modify the differential rate formula.
\begin{itemize}
\item \textbf{Atomic states to free state transitions  for deposited energy $E_{e} > 60$  eV using atomic form factor. }

For a scattering event with high energy transfer, we can safely ignore the crystal structure in Si and Ge and treat the transitions to be occurring from some atomic state to free electronic state. This approximation is indeed more reasonable for high energy transitions from core states to free state since the electronic states further away from the band gap are not substantially modified by the crystal structure. There is a slight inaccuracy though when we treat the valence states as atomic states (4s, 4p for Ge and 3s, 3p for Si). But at large for high energy transfers we do not expect a sizeable deviation from this approximation if one is to perform an exact valence to conduction calculation instead of a semi-analytical valence atomic state to free state calculation. In fact, we observe that the valence atomic state to free calculation gives a decent continuation of the valence to conduction rate spectra to energies above 60 eV.

We use the \texttt{DarkARC} code \cite{catena_atomic_2020,timon_emken_2019_3581334} to evaluate the atomic form factors. \texttt{DarkARC} uses Roothan Hartree Fock (RHF) approximation for the ground state wave functions and calculates the form factor for transition to the free state. The RHF Slater Type Orbitals (STO) coefficients for Si and Ge tabulated in \cite{bunge1993} are used as inputs to \texttt{DarkARC} to get the atomic ionization form factors for Si and Ge.
The dimensionless atomic ionization form factor evaluated in \texttt{DarkARC} as described in \cite{catena_atomic_2020} is expressed as
\begin{equation}
\left|f_{\text {ion }}^{n \ell}\left(k^{\prime}, q\right)\right|^{2}=V \frac{4 k^{\prime3}}{(2 \pi)^{3}} \sum_{\ell^{\prime}=0}^{\infty} \sum_{m=-\ell}^{\ell} \sum_{m^{\prime}=-\ell^{\prime}}^{\ell^{\prime}}\left|\int \frac{\mathrm{d}^{3} k}{(2 \pi)^{3}} \psi_{2}^{*}(\mathbf{k}+\mathbf{q}) \psi_{1}(\mathbf{k})\right|^{2}~.~\,
\end{equation}

Incorporating the form factor the ionization rate can be expressed as
\be \frac{d\mathcal{R}_{a}}{d \ln E_e}= \frac{N_T}{4} \int d E_\nu \Phi(E_\nu) \int dq \left( \frac{d \sigma}{dq} \right) \left| f_{\text{ion}}^{n,l} (q, E_e) \right|^2  ~.~\,\ee

\item  \textbf{Valence band to conduction band transtions  for deposited Energy $E_{e}\leq60$  eV using valance-to-conduction band crystal form factor. }

To evaluate the rate more reliably at low energies ($\lesssim 60$  eV), the alteration of atomic orbitals near the band gap must be taken into account. These states are computed numerically using DFT methods and has been included in calculations of previous works for Dark Matter-electron scattering.\cite{Essig:2015cda,griffin_extended_2021}

For this work, we use a modification of the \texttt{EXCEED-DM} code \cite{tanner_trickle_2022_6097642} to get the valence to conduction crystal form factor binned in magnitude of momentum transfer $q$ and deposited energy $E_{e}$ and employ it in our neutrino electron scattering rate calculations for Si and Ge. Our definition of the form factor that we extract from \texttt{EXCEED-DM} is defined as follows

\begin{eqnarray}
     \left|f_{v \rightarrow c}\left(q, E_{e}\right)\right|^{2}&=&\frac{4 \pi^{2} \Omega}{q^{3}}\sum_{i,f} \int_{\mathrm{BZ}} \frac{d^{3} k_{i} d^{3} k_{f}}{(2 \pi)^{6}} E_{e} \delta\left(E_{e}-E_{f \mathbf{k}_{f}}+E_{i \mathbf{k}_{i}}\right) \sum_{\mathbf{G}} q \delta\left(q-\left|\mathbf{k}_{f}+\mathbf{G}-\mathbf{k}_{i}\right|\right) \nonumber\\ &~& \times \left|\frac{1}{\Omega} \int_{\text {cell }} d^{3} x e^{i \mathbf{G} \cdot \mathbf{x}} u_{f, \mathbf{k}_{f}}^{*}(\mathbf{x}) u_{i, \mathbf{k}_{i}}(\mathbf{x})\right|^{2}~,~
\end{eqnarray}
where $\Omega$ is the primitive cell volume; $i(f)$ are band numbers which index the initial and final states ;
$k_i$ and $k_f$ are the initial and final Bloch momentum in the first Brillouin zone;
$\mathbf{G}$ is a reciprocal lattice vector
\begin{equation}
\begin{aligned}
\frac{d \mathcal{R}_{v \rightarrow c}}{d \ln E_{e}}=N_{cell} \int d E_{\nu} \Phi\left(E_{\nu}\right) \int dq \frac{d \sigma_{e}}{d q}&\left|f_{v \rightarrow c}\left(q, E_{e}\right)\right|^{2}
\end{aligned}
\end{equation}
where $N_{cell}$ is the number of primitive unit cells in the crystal;
$\Phi\left(E_{\nu}\right)$ is the incoming neutrino flux;
 $\frac{d \sigma_{e}}{d q}$ is the differential cross section as a function of momentum transfer assuming that the initial electron was at rest.

The DFT-computed wavefunctions with all-electron reconstruction used in the calculations are taken from \cite{sinead_m_griffin_2022_6015637}. The calculation includes 4 valence bands for Si and 4 valence bands $+$ 10 outer core bands(corresponding to 3d shell electrons) for Ge. All bands up to 60 eV above the band gap are included in the conduction band. This choice therefore covers all transitions with deposited energy below 60 eV.
The lowest energy that can be deposited is the energy band gap between occupied valence bands and unoccupied conduction bands which is 1.11 eV for Si and 0.67 eV for Ge.

As a matter of fact, the \texttt{EXCEED-DM} code provides a robust idea for calculation of scattering rate in semiconductor crystals by using a combination of Density Functional Theory (DFT) and semi-analytic methods. The work divides the electronic states in a (pure) crystal into four categories: core, valence, conduction, and free. The valence and conduction electronic band structures and wave functions are approximated by numerical calculations using DFT. The states further away from the band gap (the core and free states) are modelled using semi-analytic approximations. The total transition rate is then expressed as the sum of four contributions: valence to conduction (v to c), core to conduction(c to c), valence to free ( v to f) and core to free (c to f) transition rates. 
Though the form factors obtained from \texttt{EXCEED-DM} works well for dark matter scattering, we observe that the core to conduction (c to c), valence to free (v to f), and core to free (c to f) calculations lead to some unwarranted results for neutrino induced scattering via light mediators. On closer inspection, we find that the modelling of the initial and final states in \texttt{EXCEED-DM} leads to some fallacious behaviour in the form factor at low momentum transfers. Since the scattering rates for neutrino electron scattering via light mediator NSI are heavily dependent on contributions from low momentum transfers, we therefore only choose to use valence to conduction (v to c) form factors from \texttt{EXCEED-DM} for this work and instead approximate the remaining rate spectra for transitions to states with energies above $60$ eV using \texttt{DarkARC}. We find that this approximation using \texttt{DarkARC} in fact gives a reasonable continuation to the valence to conduction rate above $E_{e} > 60$ eV. Also, the sensitivity calculations are largely insensitive to this choice of approximating the rate using atomic form factor.
\end{itemize}

 \begin{figure}[tbp]
\centering 
\includegraphics[width=.48\textwidth]{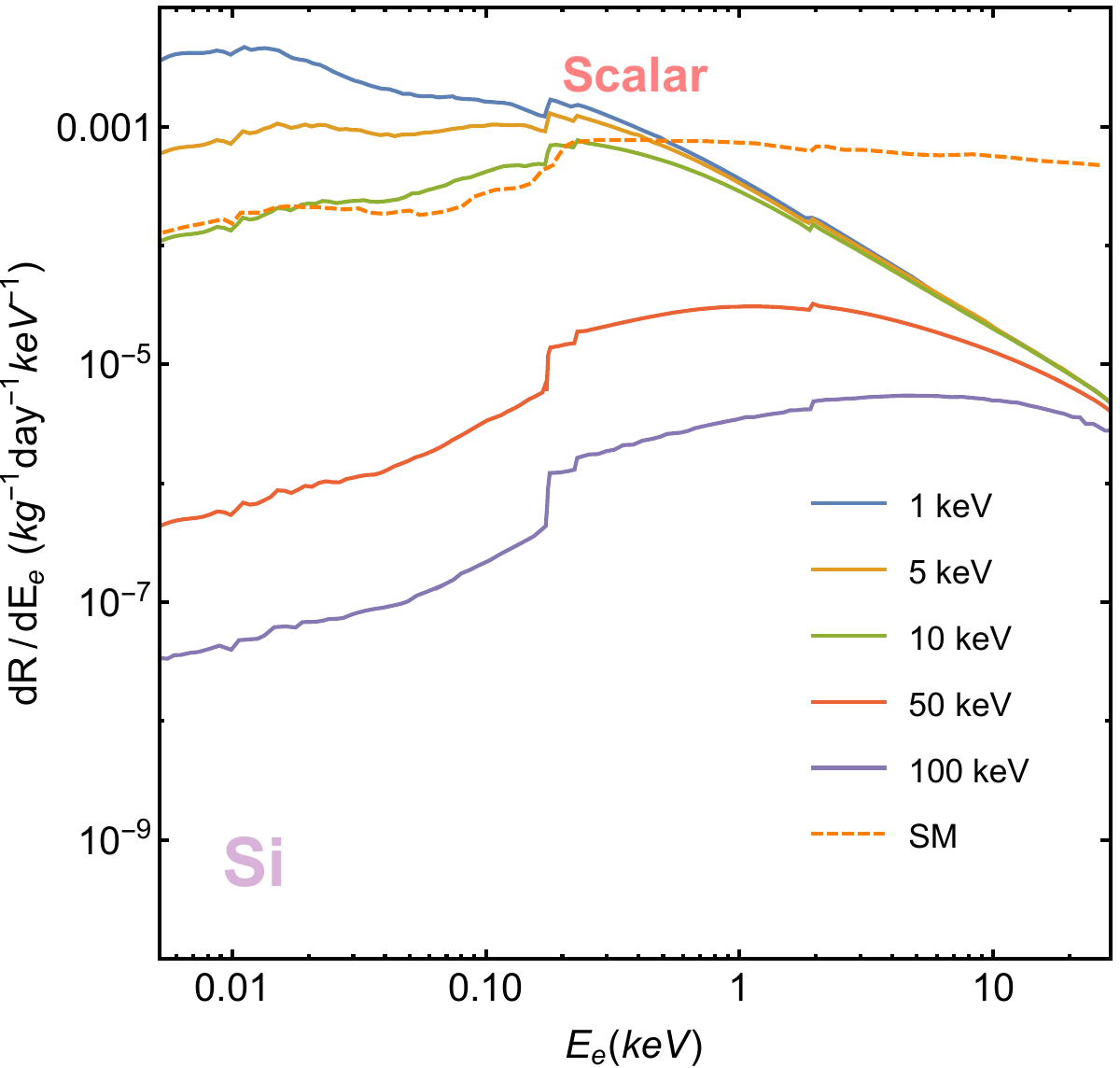}
\hfill
\includegraphics[width=.48\textwidth]{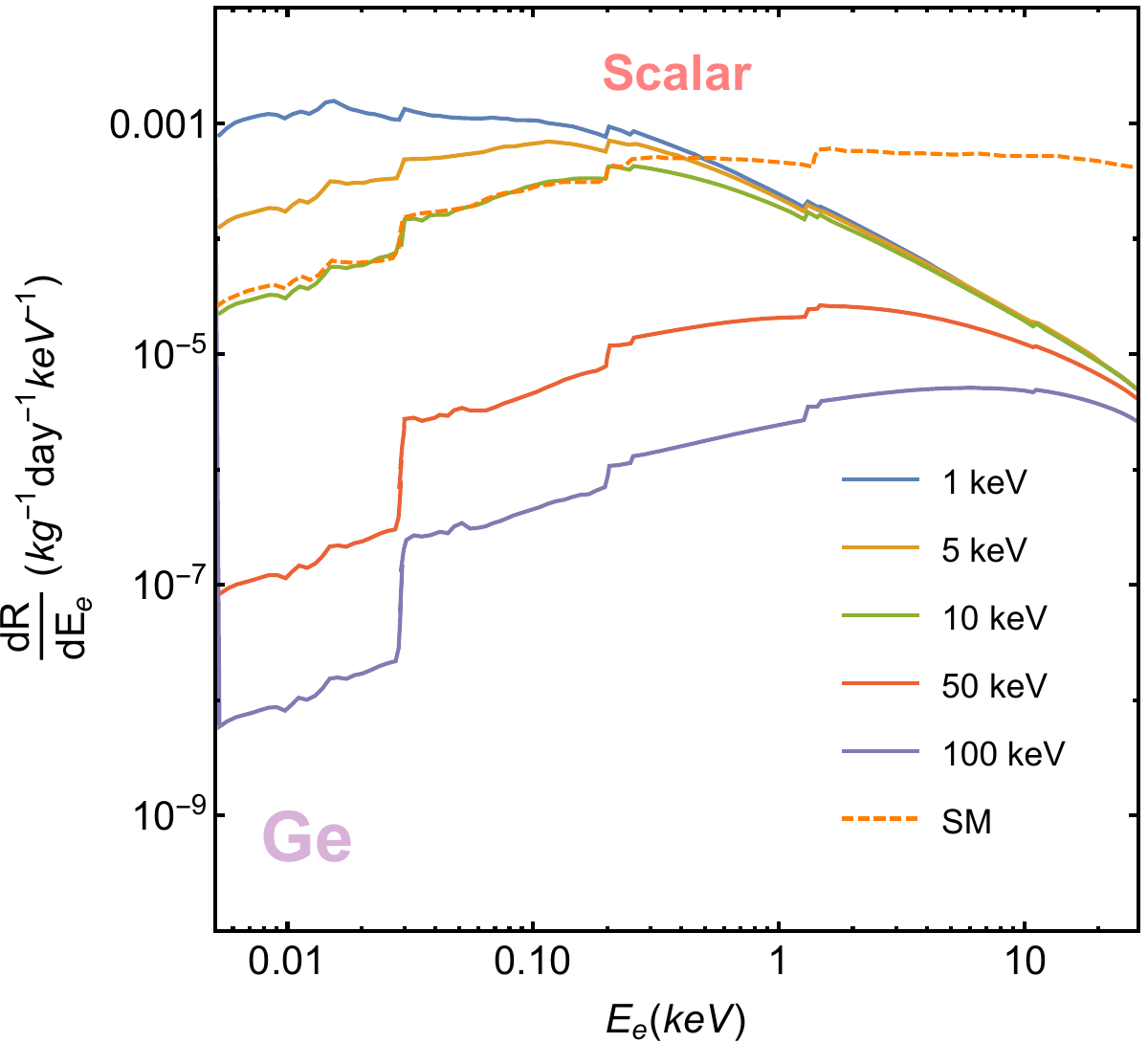}
\hfil
\includegraphics[width=.48\textwidth]{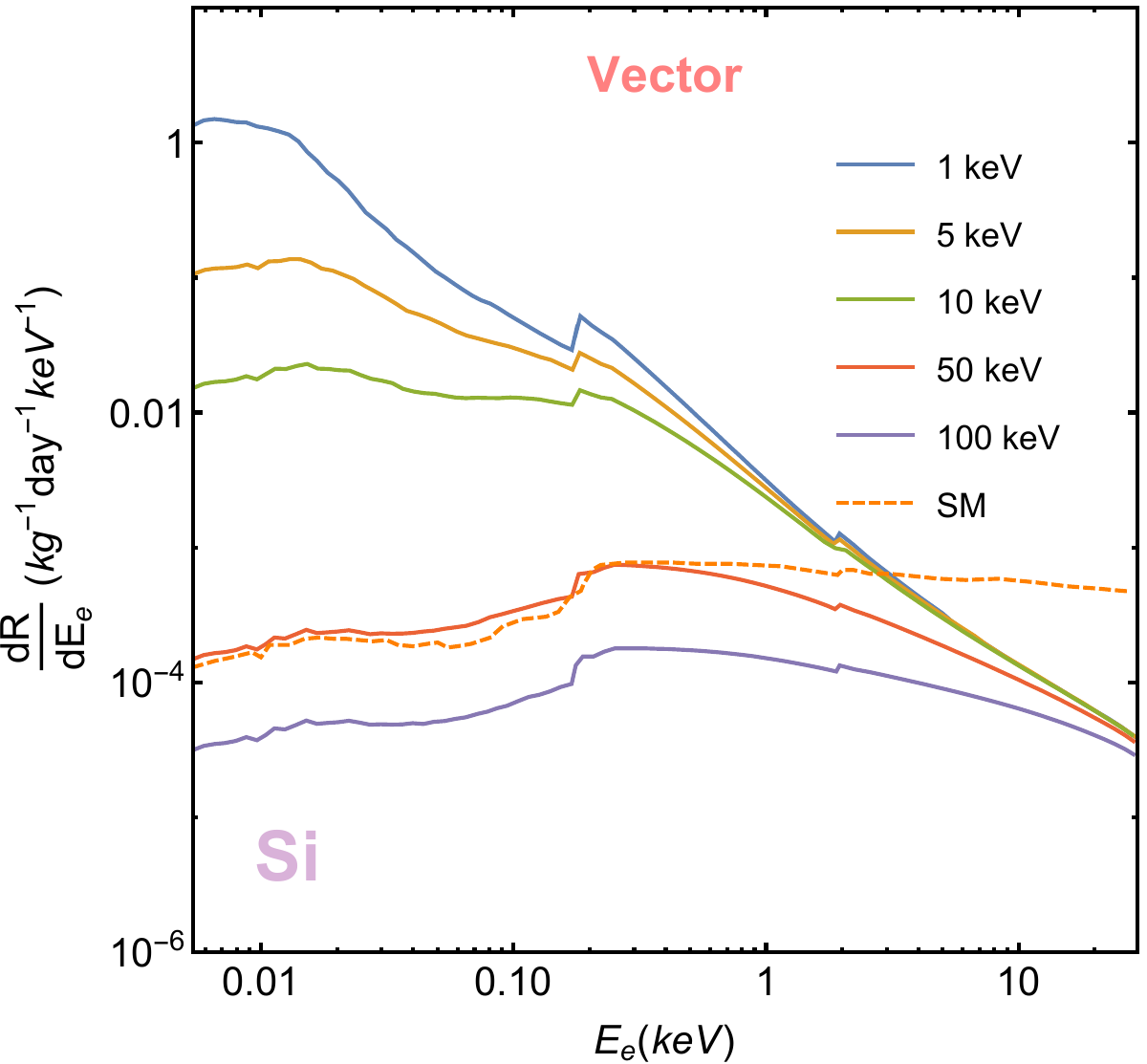}
\hfill
\includegraphics[width=.48\textwidth]{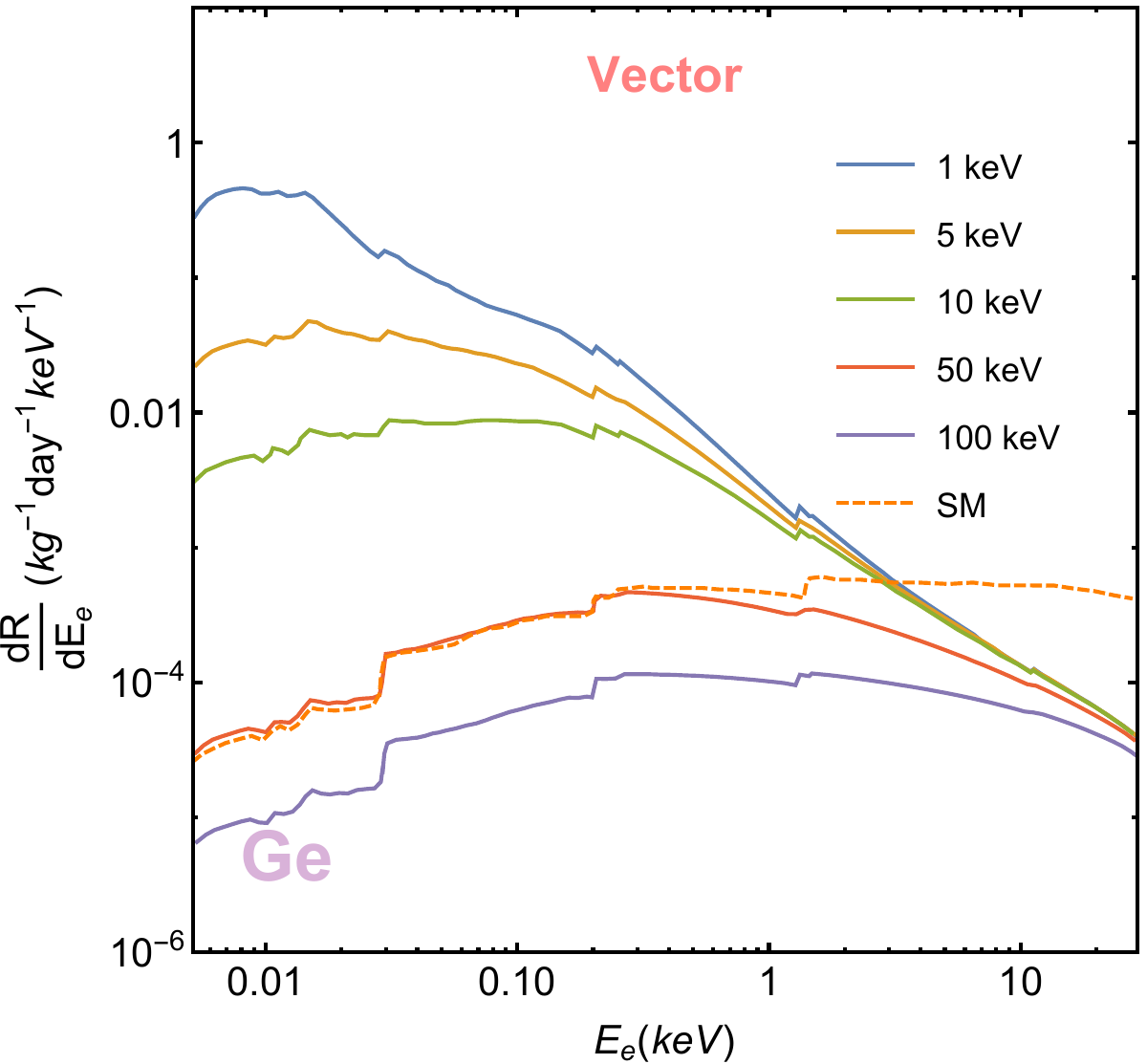}

\captionsetup{justification   = RaggedRight,
             labelfont = bf}
\caption{\label{fig:dRdEe} 
Differential event rate as a function of the deposited energy $E_e$ for neutrino scattering off silicon and germanium target for various mediator masses. The appropriate atomic and crystal form factors have been considered. The curves for the low mass mediators rise at low deposited energy. The benchmark couplings taken are: $g_{\nu}=g_{e}=5 \times 10^{-7}$ for scalar mediator (Top Panel), and
$g_{\nu}=g_{e}=1.5 \times 10^{-7}$ for vector mediator (Bottom Panel).}
\end{figure}

Recoil spectra for silicon and germanium detectors incorporating the above form factors are shown in figure~\ref{fig:dRdEe} for a few different mediator masses for both scalar and vector cases. The $x$-axis in the plots corresponds to the total energy deposited in the detector. The benchmark neutrino and electron couplings have been taken to be $g_{\nu}=g_{e}=1.5 \times 10^{-7}$ and $g_{\nu}=g_{e}=5 \times 10^{-7}$ for vector and scalar mediators, respectively. Note that,  the spectra corresponding to mediators with mass $\mathcal{O}(1)$~keV or below rise at low recoil energies. Therefore, more events would be registered in such scenarios. Thus, the low threshold detectors with very small background are ideal to probe  light mediators with masses $\le 1$~keV. 

Note that we present the recoil spectra as a function of the deposited energy. But in reality, the experiments do not measure the deposited energy directly. Rather they measure the number of electron-hole pairs produced in a scattering event, {\it i.e.}, the ionization signal, $Q$. When an incoming neutrino scatters off an electron inside the atom, it deposits energy $E_{e}$ to the electron, causing the electron to jump from the valence bands or the core states to  conduction bands. The electron produced in the primary scattering then goes through a chain of secondary scattering processes and rapidly redistributes the initial deposited energy by creating many low energy electron-hole pairs, until all the electrons and holes have energies below the pair creation threshold. A realistic modelling of the secondary scattering processes is very complicated. Therefore, a simple linear approximation can be adopted which does a decent job. Under this approximation, the ionization signal Q is related to the deposited energy $E_e$ by the expression \begin{equation} Q=1+\left\lfloor\frac{E_{e}-E_{\mathrm{g}}}{\varepsilon}\right\rfloor~,~ \end{equation} where $\lfloor~ \rfloor$ is the greatest integer function; $\varepsilon$ is the mean energy per electron-hole pair; and $E_g$ is the bang-gap energy. The values are given by: $\varepsilon$ for silicon (germanium) is 3.6 eV (2.9 eV); and $E_g$ is 1.1 eV (0.67 eV) for silicon (germanium)~\cite{1968JAP....39.2029K}.

 \section{Light Mediator Models} \label{sec:models}

 In this section we mention a few specific low-energy models with neutrino-electron interactions through the exchange of light mediators, where our analysis could be applicable. When considering the models, we confine ourselves to those with  light scalar and vector mediators only. The masses of the mediators are in the sub-MeV range. Our analysis of the neutrino-electron scattering can either put a tight bound on the parameter space of these models or can predict their discovery potential in the future and upcoming experiments.

 \subsection{Scalar mediator models} \label{models:scalar}
 
The SM scalar sector is the simplest possible scalar structure with only one Higgs doublet~\cite{Higgs:1964pj, Higgs:1966ev, Englert:1964et, Guralnik:1964eu, Kibble:1967sv}. Several well motivated extensions of the SM scalar sector have been considered in the literature. Among them, the   two Higgs doublet model (2HDM)~\cite{Lee:1973iz, Branco:2011iw} is simple and has considerable phenomenological potential. Recently, it was shown that one can generate a physical scalar spectrum with one light scalar with mass $ \mathcal{O}(100)$ MeV by extending the 2HDM with a singlet scalar field without any fine-tuning~\cite{Dutta:2020scq}. The $ \mathcal{O}(100)$ MeV scalar field can be useful to study NSI and anomalous magnetic moment problems along with other well known tensions in the experimental results  (e.g. MiniBooNE anomalous events).  But we need even a lighter scalar with mass $ \mathcal{O}$(keV) to do the analysis of neutrino-electron scattering at the reactor neutrino experiments. One needs to do some level of fine-tuning to get a scalar of mass $ \mathcal{O}$(keV) from SM symmetry only by extending the scalar sector. There are not any specific models in the literature with such a light scalar particle that generates considerable non-standard interactions of neutrinos. Therefore we propose one such model below. We briefly describe the model here in the context of our analysis. The details of the complete model and its phenomenology will appear in a future work~\cite{Dutta:2022future}. Note that, one can avoid the fine-tuning by considering a new $U(1)$ gauge symmetry or some other new physics.

 In the following, we briefly mention our new model and explain how it can generate a non-standard neutrino-scalar and electron-scalar interaction terms. We extend the scalar sector of the SM by adding one additional doublet, one scalar singlet, and one scalar triplet field. The scalar fields under $SU(2)_L\times U(1)_Y$ symmetry are:
\bea
H_1&\sim& (2,1/2),~~~~~H_2\sim (2,1/2),~~~~~\nonumber\\ \phi&\sim& (1,0),~~~~~  
\Delta \sim (3,1)~.~\,
\eea
 For simplicity, we consider that the scalar potential is CP-conserving. In general, they can be CP-violating. Also, note that we do not impose any discrete symmetry in the scalar sector. The most general renormalizable  scalar potential is then given by

\bea \label{scalarpotential}
V &=& m_1^2 H_1^\dagger H_1 + m_2^2 H_2^\dagger H_2+m_{\phi}^2 \phi^\dagger \phi+m_{\phi^\prime}^2 (\phi^2 +\phi^{\dagger 2})+m_\Delta^2 \mbox{Tr}~ \Delta^\dagger \Delta	+ m_{12}^2 (H_1^\dagger H_2 + H_2^\dagger H_1) \nonumber\\ &~&+ m_{11\phi} H_1^\dagger H_1 (\phi^\dagger+\phi)  + m_{22\phi} H_2^\dagger H_2 (\phi^\dagger+\phi)+ m_{12\phi} (H_1^\dagger H_2+H_2^\dagger H_1)(\phi^\dagger+\phi)\nonumber\\ &~& +m_{\Delta \Delta \phi}(\phi^\dagger+\phi) \mbox{Tr}~ \Delta^\dagger \Delta + m_{3\phi}(\phi^3 +{\phi^\dagger}^3) + m_{3\phi}^\prime(\phi^2\phi^\dagger+{\phi^\dagger}^2 \phi)\nonumber\\ &~&+m_{11 \Delta}(H_1^T i\sigma_2 \Delta^\dagger H_1 - H_1^\dagger \Delta i \sigma_2 H_1^*)+ m_{22\Delta}(H_2^T i\sigma_2 \Delta^\dagger H_2 - H_2^\dagger \Delta i \sigma_2 H_2^*) \nonumber\\ &~&+ m_{12\Delta}(H_1^T i\sigma_2 \Delta^\dagger H_2 - H_2^\dagger \Delta i \sigma_2 H_1^*)+m_{21\Delta}(H_2^T i\sigma_2 \Delta^\dagger H_1 - H_1^\dagger \Delta i \sigma_2 H_2^*)\nonumber\\ &~& + \frac{\lambda_1}{2} (H_1^\dagger H_1)^2 +\frac{\lambda_2}{2} (H_2^\dagger H_2)^2+\frac{\lambda_\phi}{2} (\phi^\dagger \phi)^2 +\frac{\lambda_\phi^\prime}{2} ({\phi^\dagger}^3 \phi+\phi^3 \phi^\dagger)+\frac{\lambda_\phi^{\prime\prime}}{2} ({\phi^\dagger}^4 + \phi^4) +\frac{\lambda_\Delta}{2} ( \mbox{Tr}~ \Delta^\dagger \Delta)^2 \nonumber\\ &~&+\frac{\lambda^\prime_{\Delta}}{2}\mbox{Tr}~(\Delta^\dagger \Delta)^2  + \lambda_3 (H_1^\dagger H_1)(H_2^\dagger H_2) + \lambda_4 (H_1^\dagger H_2 )(H_2^\dagger H_1) +\frac{\lambda_5}{2}\left[ (H_1^\dagger H_2)^2+(H_2^\dagger H_1)^2 \right] \nonumber\\ 
 &~&+ \lambda_6 \left[ (H_1^\dagger H_1)(H_1^\dagger H_2)+(H_1^\dagger H_1)(H_2^\dagger H_1)\right] + \lambda_7 \left[ (H_2^\dagger H_2)(H_1^\dagger H_2)+(H_2^\dagger H_2)(H_2^\dagger H_1)\right]\nonumber\\ 
&~&+\lambda_8(H_1^\dagger H_1)(\phi^\dagger \phi)+\lambda_8^\prime(H_1^\dagger H_1)({\phi^\dagger}^2+ \phi^2)+\lambda_9(H_2^\dagger H_2)(\phi^\dagger \phi)+\lambda_9^\prime(H_2^\dagger H_2)({\phi^\dagger}^2 +\phi^2) \nonumber\\ 
 &~& +\lambda_{10}(H_1^\dagger H_2+H_2^\dagger H_1)\phi^\dagger \phi +\lambda_{10}^\prime(H_1^\dagger H_2+H_2^\dagger H_1)({\phi^\dagger}^2 +\phi^2)+\lambda_{11} H_1^\dagger \Delta \Delta^\dagger H_1 \nonumber\\ 
 &~&+ \lambda_{12}(H_1^\dagger H_1)\mbox{Tr}( \Delta^\dagger \Delta)+\lambda_{13} H_2^\dagger \Delta \Delta^\dagger H_2 + \lambda_{14}(H_2^\dagger H_2)\mbox{Tr}( \Delta^\dagger \Delta)+\lambda_{15} \left[ H_1^\dagger \Delta \Delta^\dagger H_2+H_2^\dagger \Delta \Delta^\dagger H_1 \right]\nonumber\\ 
 &~& +\lambda_{16} \left[(H_1^\dagger H_2)\mbox{Tr~}  \Delta^\dagger \Delta +  (H_2^\dagger H_1)\mbox{Tr~} \Delta^\dagger \Delta \right]+\lambda_{17} (\phi^\dagger \phi)\mbox{Tr}( \Delta^\dagger \Delta)+\lambda_{17}^\prime({\phi^\dagger}^2 +\phi^2) \mbox{Tr}( \Delta^\dagger \Delta) \nonumber\\ &~&+\lambda_{18}(H_1^T i\sigma_2 \Delta^\dagger H_1  -  H_1^\dagger \Delta i \sigma_2 H_1^*)(\phi^\dagger+\phi) + \lambda_{19}(H_2^T i\sigma_2 \Delta^\dagger H_2 - H_2^\dagger \Delta i \sigma_2 H_2^*) (\phi^\dagger+\phi)
\nonumber\\ &~& +\lambda_{20}(H_1^T i\sigma_2 \Delta^\dagger H_2  - H_2^\dagger \Delta i \sigma_2 H_1^*)(\phi^\dagger+\phi)+\lambda_{21}(H_2^T i\sigma_2 \Delta^\dagger H_1 - H_1^\dagger \Delta i \sigma_2 H_2^*)(\phi^\dagger+\phi)~.~\,
\eea
 
 By tuning the parameters of the scalar potential, we can achieve a scenario where only one of the doublet $H_1$ obtains vacuum expectation value (VEV), $\braket{H_1}=v_1/\sqrt{2}$, which completely controls the spontaneous electroweak symmetry breaking. The triplet field $\Delta$ and the singlet field $\phi$ also obtain VEVs in the process. The VEV of $\Delta$ is constrained by the electroweak precision measurement of the $\rho$ parameter~\cite{Langacker:1980js, ParticleDataGroup:2020ssz}, given by $v_{\Delta} < 0.032 v_1$, where $v_1 =246$~GeV. The vev of the $\phi$ field is unconstrained and we choose to work with $1$~ GeV value.
 
 After the spontaneous symmetry breaking,  the scalars can be expressed as
\begin{eqnarray}
	 &H_1 &\sim   \left( \begin{array}{c}  {G}^+  \\  \frac{1}{\sqrt{2}}(v_1 +\rho_1+iG_0)  \end{array} \right),~~~ H_2 \sim   \left( \begin{array}{c}  {\phi_2}^+ \nonumber \\  \frac{1}{\sqrt{2}}(v_2 +\rho_2+i\eta_2)  \end{array} \right),~~~~\\&\phi&\sim \frac{1}{\sqrt{2}}(v_\phi+\rho_\phi+i\eta_\phi),~~~ \Delta \sim    \left( \begin{array}{cc} \Phi^+ & \sqrt{2}\Phi^{++} \\ \frac{1}{\sqrt{2}}(v_\Delta +\rho_t+i\eta_t) & -\Phi^+ \end{array} \right)   ~.~\,
\end{eqnarray}
 
 There are total 16 scalar degrees of freedom (dof) present in the model. Out of that 3 dof ($G^\pm$ and $G^0$) become the longitudinal polarization states of the vector gauge bosons $W^{\pm}$ and $Z$ and make them massive. The rest of them become physical scalar particles. Two doubly charged physical scalars $h_3^{\pm \pm}$ come from $\Phi^{\pm \pm}$ while 4 single charged physical scalars, named $h_1^\pm$ and $h_2^\pm$ come from the mixing of $\phi_2^\pm$ and $\Phi^\pm$. There are total 4 CP-even states $\rho_1$, $\rho_2$, $\rho_\phi$ and $\rho_t$. They mix among themselves and generate four physical neutral Higgs particles named as $h$, $h_1$, $h_2$ and $h_3$. We identify $h$ as the SM Higgs particle having mass $125.5$~GeV. Here, $h_1$ can be a  very light scalar with mass $\sim \mathcal{O}(1)$~keV. The three CP-odd states $\eta_2, \eta_\phi$ and $\eta_t$ mix and give three physical neutral pseudoscalars, we call them $s_1, s_2$ and $s_3$. All of them have masses that satisfy the LHC bounds.
 
 \begin{table}[tbp]
\centering
\begin{tabular}{c|lr}
\hline
Parameters & Descriptions &~~~ Benchmark values\\
&&\\

\hline 
\makecell{$m_1^2$, $m_2^2$, $m_{\phi}^2$, $m_{\phi^\prime}^2$, \\ $m_{\Delta}^2$ and $m_{12}^2$} & \makecell{these are the mass square  \\ parameters  $\sim [\mathcal{O}$(100) GeV]$^2$}  &~~~ \makecell{  $m_{1}^2 = -(126)^2$~GeV$^2$, \\ $m_{2}^2 = (843)^2$~GeV$^2$, \\ $m_{\phi}^2 = m_{\phi}^{\prime 2}  = -(54)^2$~GeV$^2$, \\ $m_{\Delta}^2 = (703)^2$~GeV$^2$, \\ $m_{12}^2 = - (100)^2$~GeV$^2$. }\\

&& \\

\makecell{ $m_{11\phi}$, $m_{22\phi}$, $m_{12\phi}$, $m_{\Delta \Delta \phi}$, \\ $m_{3\phi}$, $m_{3\phi}^\prime$, $m_{11\Delta}$, $m_{22\Delta}$, \\ $m_{12\Delta}$ and $m_{21\Delta}$ } & \makecell{the mass dimension of \\ these parameters are one \\ with values $\sim \mathcal{O}(0.1-10)$~GeV} &~~~ \makecell{$m_{11\phi} = 0.138$~GeV, \\ $m_{11\Delta} = 11.54$~GeV \\ and rest of them \\ are $-10$~GeV. }  \\

&& \\

\makecell{ $\lambda_{1}$, $\lambda_{2}$, $\lambda_{\phi}$, $\lambda_{\phi}^\prime$, $\lambda_{\phi}^{\prime}$  $\lambda_{\Delta}$ \\ $\lambda_{\Delta}^\prime$,  $\lambda_{3}$, $\lambda_{4}$, $\lambda_{5}$,  $\lambda_{6}$, $\lambda_{7}$, \\ $\lambda_{8}$, $\lambda_{8}^\prime$, $\lambda_{9}$, $\lambda_{9}^\prime$, $\lambda_{10}$, $\lambda_{10}^\prime$, \\ $\lambda_{11}$, $\lambda_{12}$, $\lambda_{13}$, $\lambda_{14}$, $\lambda_{15}$, \\ $\lambda_{16}$,   $\lambda_{17}$, $\lambda_{17}^\prime$, $\lambda_{18}$, $\lambda_{19}$, \\ $\lambda_{20}$ and $\lambda_{21}$ } & \makecell{these are the dimensionless \\ couplings with values \\ between $ 0.1-1.0$} &~~~ \makecell{ $\lambda_{1} = 0.53$, $\lambda_{6} = 0.33$;  \\  rest of them are 0.1 }\\
\hline
\end{tabular}
\captionsetup{justification   = RaggedRight,
             labelfont = bf}
\caption{\label{tab:scalarparameters} Brief descriptions and the necessary range of values for all the parameters that appear in the scalar potential in eq.~\ref{scalarpotential}. One particular benchmark scenario is shown. We use this benchmark to generate a light scalar of mass $\sim \mathcal{O}(1)$~keV along with other heavy physical scalars consistent with the LHC bounds. The values of the vev's are $v_1 = 246$~GeV, $v_2 = 0 $~GeV, $v_\phi =  1$~GeV and $v_\Delta = 1 $~GeV.  }
\end{table}

The Yukawa sector Lagrangian in the interaction basis is given by \bea \label{yukawa} -\mathcal{L}_{Yukawa} &=& \bar{q}^\prime_{L_i}(y^\prime_{1d})_{ij} d^\prime_{R_j} H_1 +\bar{q}^\prime_{L_i}(y^\prime_{1u})_{ij} u^\prime_{R_j} \tilde{H_1}+\bar{l}^\prime_{L_i}(y^\prime_{1e})_{ij} e^\prime_{R_j} H_1 + \bar{q}^\prime_{L_i}(y^\prime_{2d})_{ij} d^\prime_{R_j} H_2 \nonumber\\ &+&\bar{q}^\prime_{L_i}(y^\prime_{2u})_{ij} u^\prime_{R_j} \tilde{H_2} +\bar{l}^\prime_{L_i}(y^\prime_{2e})_{ij} e^\prime_{R_j} H_2 + \bar{l}^{\prime c}_{L_i} (y^\prime)_{ij} i \sigma_2 \Delta l^\prime_{L_j} ~,~\eea where $i,j$ are the family indices with values $1,2$, and $3$; also note that the primed fermions are the interaction states. The first three terms give rise to the masses of the charged leptons and quarks upon diagonalization. The next three terms give interaction with $H_2$ which is not constrained by the fermion masses. Note that, these interactions do not respect the SM flavor symmetries and thus can generate flavor changing neutral currents at tree level through the interactions with the neutral components of $H_2$.  The fermions have no direct interactions with the scalar singlet, the only way they can interact with the singlet is via the scalar mixing. Therefore, the physical scalars, light and heavy, interact with all flavors of fermions, diagonally and off-diagonally.

\begin{table}[tbp]
\centering
\begin{tabular}{c|l}
\hline
Physical particles & \makecell{Mass values \\ and descriptions }  \\

&\\

\hline 
\makecell{Neutral scalars:  \\$h$, $h_1$, $h_2$ and $h_3$} & \makecell{$m_h = 125.5 $~GeV, \\ this is the SM Higgs;  \\ $m_{h_1} = 0.001 $~GeV, \\very light scalar, \\neccessary for our analysis; \\ $m_{h_2} = 500 $~GeV, \\ $m_{h_3} = 600 $~GeV \\ they satisfy the LHC bounds. }  \\

&\\

\makecell{Neutral pseudo-scalars:  \\ $s_1$, $s_2$ and $s_3$} & \makecell{$m_{s_1} = 500 $~GeV, \\ $m_{s_2} = 500 $~GeV, \\$m_{s_3} = 600 $~GeV, \\ they satisfy LHC bounds.} \\

&\\

\makecell{Single charged scalars:  \\ $h_1^{\pm}$, $h_2^\pm$} & \makecell{$m_{h_1^{\pm}} = 500 $~GeV, \\ $m_{h_2^\pm} = 600 $~GeV, \\ in agreement with LHC bounds.} \\

&\\

\makecell{Double charged scalars:  \\ $h_3^{\pm \pm} $} & \makecell{$m_{h_3^{\pm \pm}} = 500 $~GeV, \\ satisfying LHC limits. } \\
\hline
\end{tabular}
\captionsetup{justification   = RaggedRight,
             labelfont = bf}
\caption{\label{tab:scalarspectrum} The physical scalar spectrum generated using the benchmark values from Table~\ref{tab:scalarparameters}. The physical scalar $h$ is identified as the SM Higgs scallar. The $h_1$ is the light scalar which will be identified as $\phi^\prime$ in the rest of the paper. The scalar masses are consistent with the LHC bounds. }
\end{table}

The last term of eq.~\ref{yukawa} is the most interesting term as it is generating the non-standard interactions of neutrinos and neutrino mass. It generates a Majorana type interaction term $- \bar{\nu}^{\prime c}_{L_i} i (y^\prime)_{ij} \nu^\prime_{L_j} \rho_t$, where $\rho_t$ is the CP-even component of the neutral component of $\Delta$ field. The $\rho_t$ mixes with other CP-even states and generates a very light physical scalar particle $h_1$. Therefore in the mass basis the term $\bar{\nu^c} \nu h_1$ is giving the non-standard neutrino-scalar interaction term.  On the other side, the light scalar $h_1$ couples to both leptons and quarks diagonally and off-diagonally. Therefore, in general, there are all possible non-standard neutrino interactions with SM fermions mediated by $h_1$. Also, note that these interactions are possible for all the other heavy physical scalars and pseudo-scalars. Therefore, the physical basis Lagrangian responsible for the non-standard neutrino interactions can be written as follows~\footnote{Note that, we only assume real Yukawa couplings. But in general they can be complex as well. If the couplings are complex then they can give rise to CP-violation in the non-standard neutrino sector~\cite{Dutta:2022future02}, which can be used to address the recently found tension between T2K and NO$\nu$A data~\cite{patrick_dunne_2020_3959558, alex_himmel_2020_3959581}.} \begin{equation}    -\mathcal{L}_{Yukawa} \supset   \bar{{\nu}^c}_i (y_{\nu})_{ij} \Gamma \nu_j h_k + \bar{f}_i (y_{f})_{ij} \Gamma f_j h_k ~,~\end{equation} where $i$ and $j$ are flavor indices, $i,j = 1,2,3$; $f= d,u,e$;  $h_k = h, h_1, h_2, h_3, s_1, s_2, s_3$; and $\Gamma = 1$ for scalars while  $\Gamma = \gamma^5$ for the pseudo-scalars. As we are only interested  in the neutrino-electron interactions, we assume that only the electron diagonal interactions are non-zero. In general, there is no strong reason to forbid the other interactions. Actually it is very interesting, though computationally very complex, to turn on all possible interactions. In that case, a global analysis based on the neutrino oscillation and scattering data will affect the parameter space. Previously such  analysis was performed mostly for the heavy mediator case~(see for example~\cite{Dutta:2020che, Coloma:2022umy, Esteban:2018ppq, Esteban:2019lfo, Khan:2016uon, Cheung:2021tmx, Denton:2022pxt}). A similar global analysis for the light scalar case will be very crucial to get bounds for a realistic light scalar model~\cite{Dutta:2022future}. From this point forward,   the light scalar $h_1$ will be identified as $\phi$. In this work, we only consider the electron and $\nu_e$ couplings with the light mediators to investigate them at the reactors.

\subsection{Vector  Models}

Here we mention a few well-studied models that can have neutrino-electron interactions through a light vector boson mediator. The extension of the SM by a $U(1)$ gauge symmetry can have the required interactions and has been well-studied.  We consider three gauge symmetries $U(1)_{L_e-L_\mu}$, $U(1)_{L_e-L_\tau}$, and $U(1)_{L_\mu-L_\tau}$~\cite{Foot:1990mn, He:1990pn, He:1991qd} that are anomaly-free without any additional particles. Note that, for the first two scenarios one can get neutrino-electron interactions at tree level. The necessary terms of the Lagrangian that can give rise to neutrino-electron couplings are, \begin{eqnarray}
     -\mathcal{L} = \bar{l}_{L_i} \gamma^\mu l_{L_i} A^\prime_\mu + \bar{e}_{R_i} \gamma^\mu e_{R_i} A^\prime_\mu - \bar{l}_{L_j} \gamma^\mu l_{L_j} A^\prime_\mu - \bar{e}_{R_j} \gamma^\mu e_{R_j} A^\prime_\mu~,~
\end{eqnarray} where $i$ and $j$ are not equal. In all three cases, the SM fermions are charged under both the new $U(1)$ gauge symmetry as well as the SM hypercharge symmetry, and thus a kinetic mixing term automatically arises between the new gauge boson $A^\prime$ and the SM hypercharge boson at one loop level even if we consider that the tree level kinetic mixing is zero. And this allows to have the desired neutrino-electron coupling in the third scenario. The coupling will be suppressed by the kinetic mixing factor which is roughly $\epsilon_{\mu \tau} \sim e g_{\mu \tau} / 4\pi^2$.

Another possible gauge group is the difference between baryon and lepton number $U(1)_{B-L}$. This group can be anomaly-free if one adds right-handed neutrinos. The kinetic mixing at one-loop level is unavoidable. We can get the neutrino-electron coupling at tree level as follows \begin{eqnarray}
    -\mathcal{L} = - \bar{l}_{L} \gamma^\mu l_{L} A^\prime_\mu + \bar{e}_{R} \gamma^\mu e_{R} A^\prime_\mu~.~\,
\end{eqnarray}

The neutrino-electron interaction is also possible in the low energy incarnation of the gauge group $U(1)_{T3R}$~\cite{Dutta:2019fxn, Dutta:2022qvn}, where only the right handed SM particles are charged. Right handed neutrino fields are needed to cancel the anomalies. The desired interaction is possible either using active-sterile neutrino mixing or using the mixing of the new gauge boson $A^'$ with SM $Z$ gauge boson.

The details of these models are not necessary for our purpose. We briefly mention various constraints on these models in the next subsection.

\subsection{Constraints}

One common feature of these models would be that they have neutrino-neutrino, neutrino-electron and electron-electron interactions induced by the new light mediators. Therefore, the parameter space of these models will be constrained by various laboratory based experiments and/or astrophysical/cosmological data.

Neutrino-electron scattering experiments such as Borexino~\cite{BOREXINO:2018ohr} and Gemma~\cite{Beda:2013mta} are quite sensitive to the modification of the scattering cross section by the non-standard interactions of neutrinos with electrons mediated by low mass mediators. The non-observation of any anomalous events at these experiments can be converted into bounds on the parameter space of coupling-mediator mass plane.  We have translated the bounds into constraints on the couplings of the light mediators (both scalar and vector) and shown them in figure~\ref{fig:sensitivitypoint1dru}. The parameter space is also constrained by the XENONnT experiment. We used the recent experimental data~\cite{XENON:2022mpc} to find the corresponding bound and have shown in figure~\ref{fig:sensitivitypoint1dru}. Similar bounds can be found in ref.~\cite{A:2022acy, Khan:2022bel}.

Different astrophysical processes such as energy loss of supernova 1987A, solar cooling process, and cooling of stars in globular clusters can also put bounds on the relevant parameter space for light mediator models. The light mediator particle can be produced inside the core of the supernova and then subsequently escape, resulting in energy loss, which can be compared to the SN1987A data to find bounds on light mediator mass and couplings. Similarly, they can be produced inside the sun or any other stars in a globular cluster, and contribute to the cooling process as well. One can derive constraints from these cooling processes by requiring that the luminosity due to the new mediator particle be small compared to the luminosity due to the photon. Note that, these astrophysical bounds can be avoided in  light mediators models by using the chameleon effect. In general, the masses of the mediators depend on the background matter density. The environment in the core of a supernova or a star is extremely dense. In such a highly dense environment, the effective mass of the mediator particle can be larger than it is on Earth, and thus the allowed parameter space can be increased. Astrophysical bounds for light scalar mediator can be found in Ref.~\cite{Babu:2019iml, Venzor:2020ova, Huang:2017egl} and the relevant bounds for vector boson were discussed in Ref.~\cite{Harnik:2012ni}.
 
The light mediators can be produced in the early Universe and can be in equilibrium with the thermal bath. Therefore, they can contribute to the $\Delta N_{eff}$ depending on their abundance during the time of recombination. If $m_{\phi, A} < 1$~MeV, then $\phi$ can decay into $\nu \nu$ final state at tree level. Note that, $\gamma \gamma$ final state is also possible for $\phi$ in this case via a one-loop diagram consisting of an electron inside the loop. The detailed analysis is out of the scope of this paper, however, we refer Ref.\cite{Escudero:2019gvw} where a detailed analysis has been performed for similar mass ranges of a light mediator.
   
  Non-standard interactions of the neutrinos can have effects on the oscillation and scattering experiments as well. Low threshold experiments based on coherent elastic neutrino-nucleus scattering (CE$\nu$NS) can place constraints on both scalar and vector non-standard interactions. Vector interaction introduces a new flavor-dependent matter potential during the propagation through matter that can affect neutrino oscillation. Current data from COHERENT can put constraints for mediator mass $\geq 10$~MeV~\cite{Denton:2018xmq}. Future reactor CE$\nu$NS experiments can probe very light mediators. The light scalar mediators, on the other hand, do not give rise to a matter potential as the scalar interaction cannot be converted into a vector current. Rather it contributes as a correction to the neutrino mass matrix. The correction factor is proportional to the inverse mass square of the mediator, therefore a small mediator mass can produce large enough interactions~\cite{Ge:2018uhz, Babu:2019iml}.

\section{Results} \label{sec:results}


In this section, we describe our strategy for the numerical analysis to probe the light scalar and vector mediated non-standard neutrino-electron scattering. Then we present the results of our analyses  for silicon and germanium detectors. Our analyses are focused towards low threshold reactor based neutrino experiments. We also give the expected total number of events as a function of the detector threshold. First, we briefly describe the key aspects of our analysis. 
\begin{itemize}
    \item The differential cross sections of neutrino-electron scattering for both scalar and vector cases increase at low recoil energies with decreasing mediator masses, as evident from figure~\ref{fig:dsigdT}. This causes a significant enhancement in the differential rate spectra at low recoils. It can be seen explicitly in figure~\ref{fig:dRdEe}. In particular, for a light vector mediator with mass $\sim 1$ keV almost 99$\%$ of the events recorded are expected to deposit less than a keV of energy in the detector. Therefore, the experiments with low threshold detectors are expected to be sensitive to light mediators with masses $\le \mathcal{O}(1)$~keV. For our analyses, we consider detectors with threshold as low as 5 eV.
    \item A major background to the neutrino-electron scattering events in a typical low threshold detector (e.g. CCD detector) comes from CE$\nu$NS interactions mostly below $\mathcal{O}(1)$~keV, since both $\nu-e$ and CE$\nu$NS have the same experimental signature in these detectors. Most low threshold detector technologies being utilized in current  experiments can not discriminate between nuclear recoil (NR) and electron recoil (ER) signals at very low recoil energies due to signal to noise limitations. Therefore, the experimental sensitivity to the non-standard neutrino-electron couplings in reactor experiments will be  dependent on the capability of ER and NR events discrimination of the detector down to very low recoil energies. Recently it was shown that a new hybrid detector at MINER  incorporating both ZIP (Z-sensitive Ionization and Phonon detector) and HV (High Voltage) technology is capable of discriminating ER and NR up to a very low threshold ($\sim 1$~keV) and would be able to do so even at a lower threshold in near future~\cite{Neog:2020ily}. We assume that the detector is capable to discriminate between NR and ER signals down to the threshold of 5 eV. Note that, above $\mathcal{O}(1)$~keV recoil energies there is no contribution from CE$\nu$NS interactions, therefore the discrimination between ER and NR is not needed.  But in this region, the rate of non-standard neutrino-electron scattering is very low and does not give any interesting result. 
    \item The main background for such analyses is the SM neutrino-electron scattering events mediated by the weak interactions. In addition to that, there would be more background events specific to the experiments. Here we assume that there exists a constant background over the SM value. We consider two such cases:  0.1 DRU and 1.0 DRU (where [DRU] = [counts $\cdot$ kg$^{-1}$ keV$^{-1}$ day$^{-1}$]). The second case is achievable at the present facilities whereas the first case can be achieved in near future~\cite{MINER:2016igy}. We perform our analyses for both cases.
\end{itemize}

We perform a $\chi ^2$ test over the distribution of deposited energy to find the bound on the parameter space in the coupling-mass plane. We start by assuming a set of independent background expectations $B_i$ and an expected signal $S_i$, which depends on the coupling and mediator mass, across bins $i = 1, ..., \nu$. If the experiment sees data consistent with the null hypothesis (only background events), the median value of the $\chi^2$ distribution tested against our $S + B$ model will exceed the $1 -\alpha $ quantile of the background only $\chi^2$ distribution. In the absence of real experimental data, we generate a set of independent, normally distributed pseudo data $d_i \sim N (B_i, \sqrt{B_i})$ for each bin.

However, in the limit of a large number of pseudo-experiments, the $1-\alpha$ quantile of this distribution approaches the theoretical limit of the inverse cumulative distribution function $F^{-1} (1-\alpha, \nu) \equiv \chi^2_\alpha$ which allows us to construct a simplifying heuristic for the separation between the two $\chi^2$ distributions described above. The expectation value of the $\chi ^2$ tested against the $S+B$ model can be written as, \begin{equation} \label{eq:chisq}  E\left[ \chi^2_{S+B} \right] = E\left[ \sum_{i=1}^{\nu} \frac{(d_i-S_i-B_i)^2}{B_i} \right] \end{equation} Using the linearity of expectation values and the assumption that $d_i$ are independent, and $Var[d_i] = B_i$, we can simplify eq.~\ref{eq:chisq}, \begin{equation} E\left[ \chi^2_{S+B} \right]  = \nu + \sum_{i=1}^{\nu} \frac{S^2_i}{B_i} \end{equation} Therefore, the test statistic can be defined as, \begin{equation} t = \sum_{i=1}^{\nu} \frac{S^2_i}{B_i} = E\left[ \chi^2_{S+B} \right] - \nu \end{equation} For relatively large degrees of freedom, Median$[\chi^2]$ $\simeq E[\chi^2]$. On the other hand, for large number of pseudo experiment and data consistent with background-only scenario, we have  $\text{Median}[\chi^2_{S+B}] >\chi^2_\alpha $. Therefore, we need to find the smallest signal such that \begin{equation}
t \gtrsim \chi^2_\alpha - \nu
\label{eq:test_stat}
\end{equation}
By using Eq.~\ref{eq:test_stat}, this method circumvents the need to generate pseudoexperiment monte carlo as a matter of computational ease while sacrificing little in the accuracy of the test statistic. The validity can be measured by generating pseudoexperiment data with the assumptions above and checking that the separation in the null-hypothesis $\chi^2$ distribution and the $S+B$ model's $\chi^2$ distribution matches well with the heuristic measure in Eq.~\ref{eq:test_stat}.
In order to get the exclusion curve at 90$\%$ confidence level, we consider $\alpha = 0.1$ and take number of bins as $\nu =37$ in the log space of the deposited energy.


 \begin{figure}[tbp]
\centering 
\includegraphics[width=.45\textwidth]{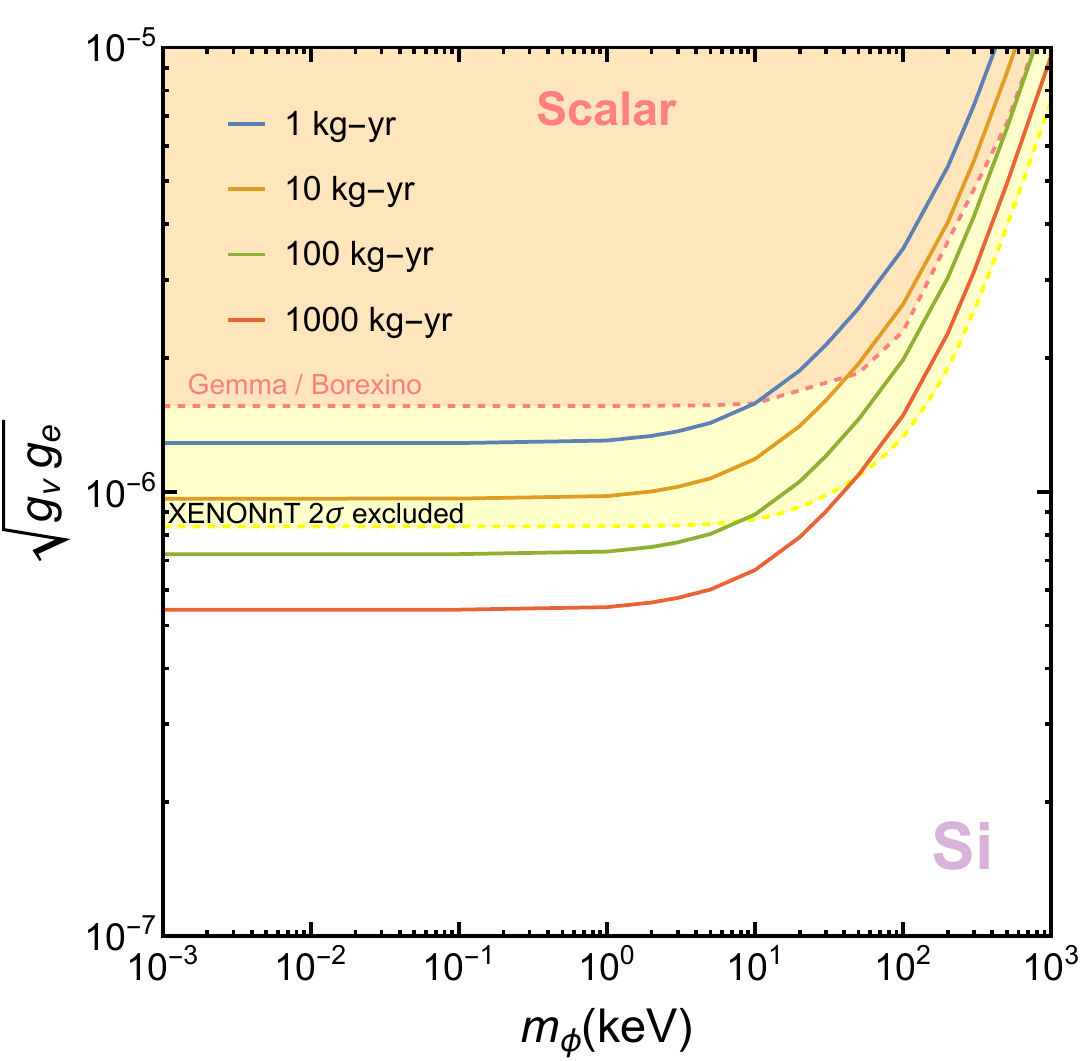}
\hfill
\includegraphics[width=.45\textwidth]{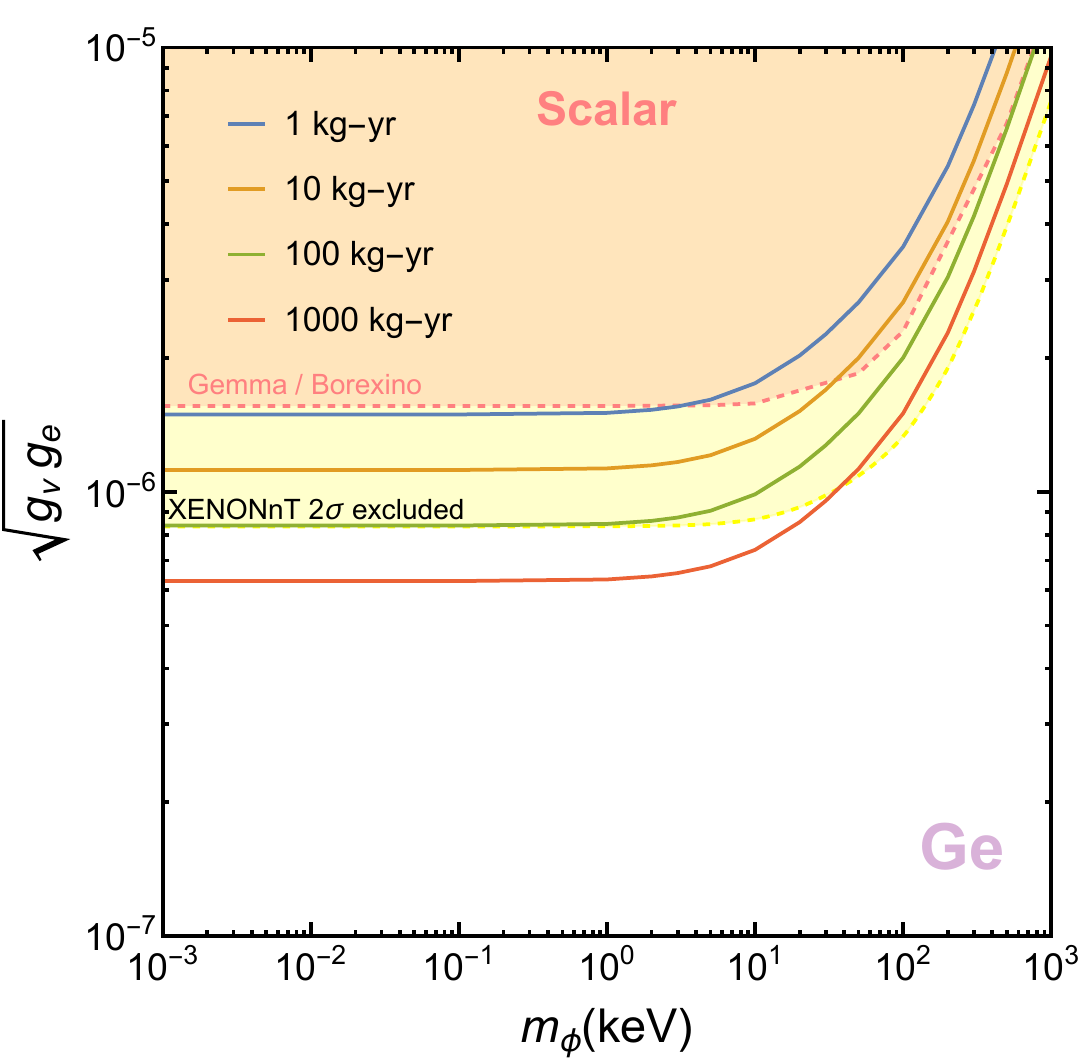}
\hfill
\includegraphics[width=.45\textwidth]{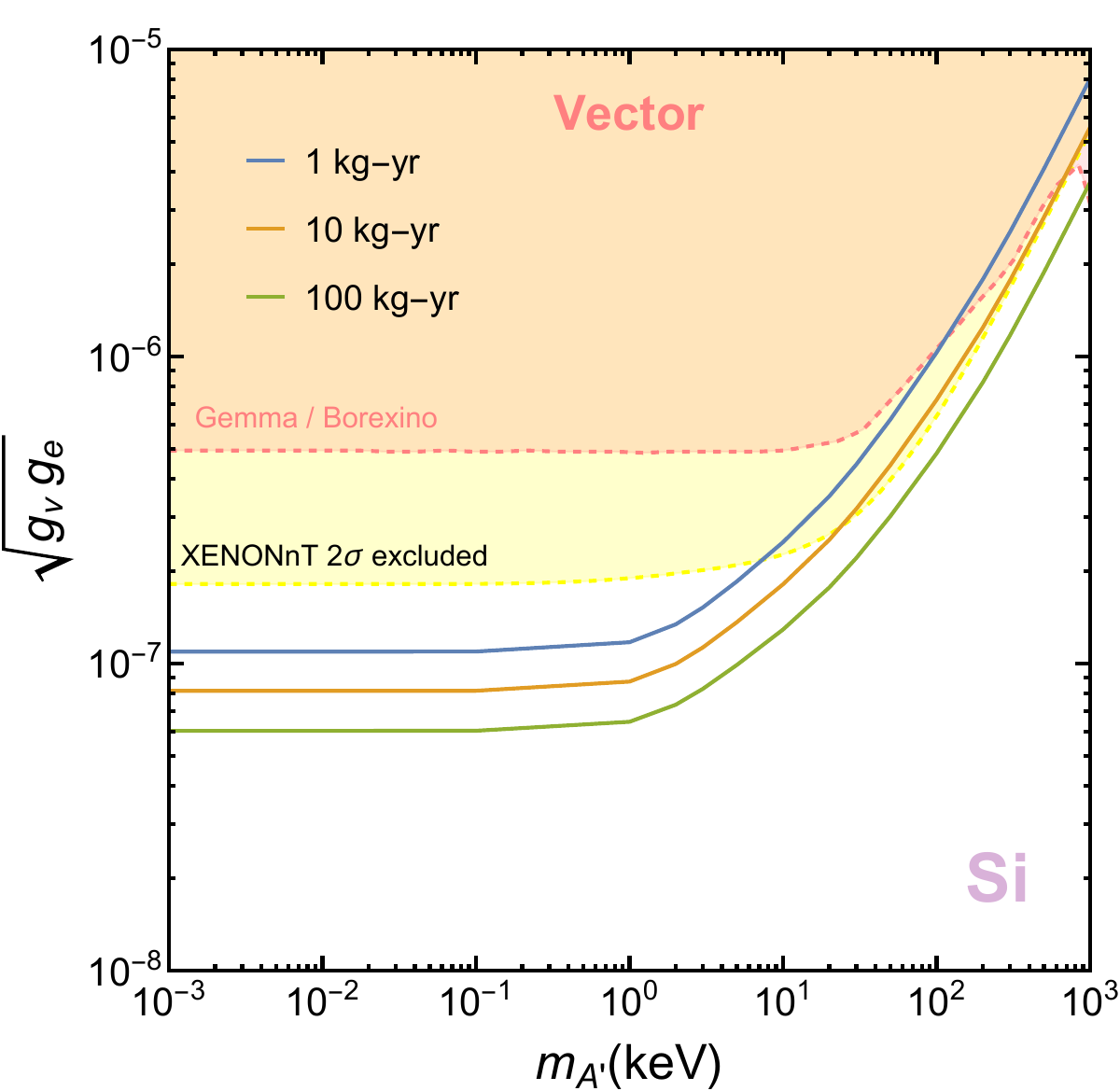}
\hfill
\includegraphics[width=.45\textwidth]{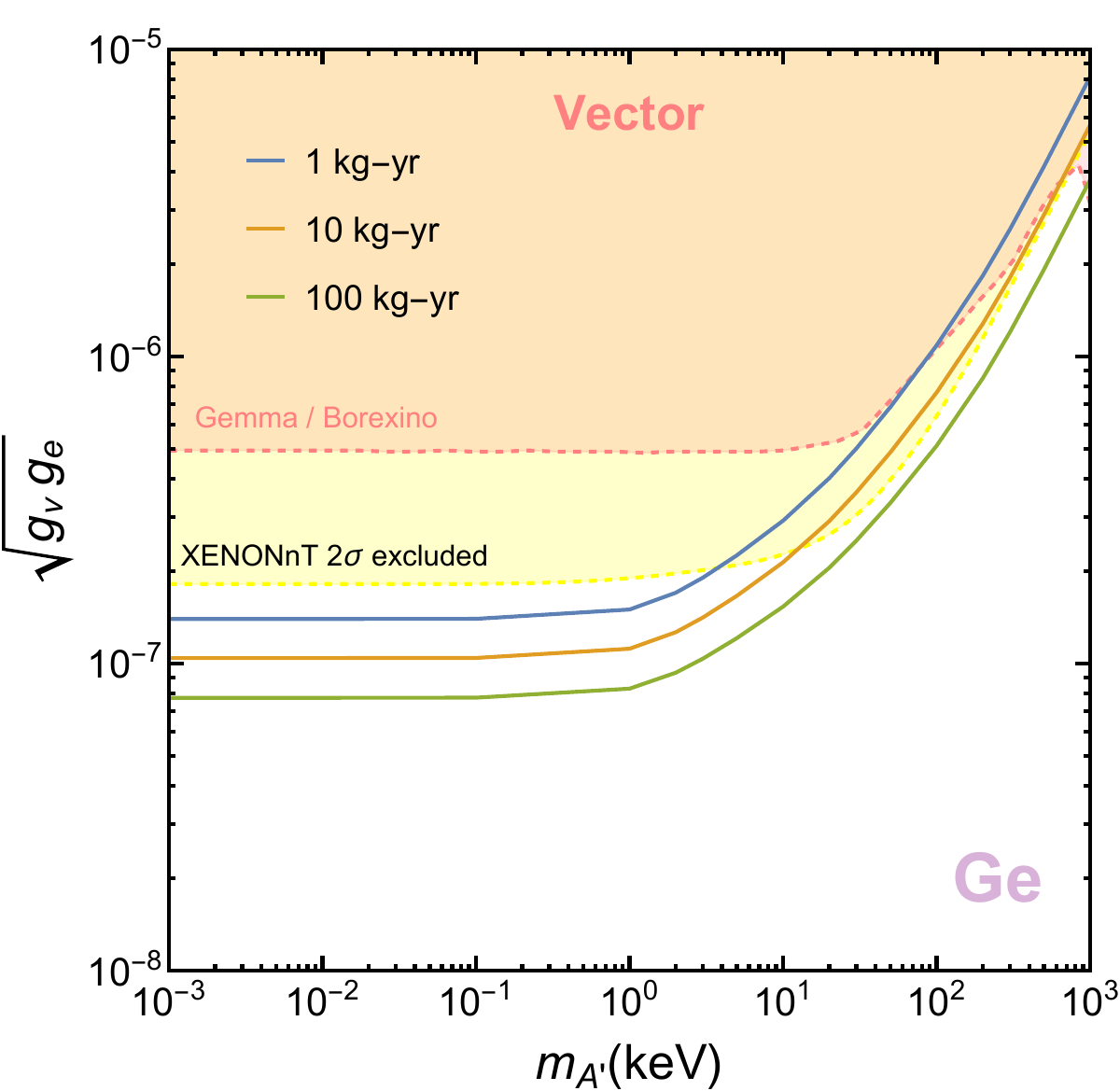}
\captionsetup{justification   = RaggedRight,
             labelfont = bf}
\caption{\label{fig:sensitivitypoint1dru}Projected bounds on the coupling $\sqrt{g_\nu g_e}$ at $90 \%$ confidence level as a function of the mediator masses. The bounds are given for four different exposure values: 1 kg-yr, 10 kg-yr, 100 kg-yr, and 1000 kg-yr for scalar mediators (top panel) for silicon (left) and germanium (right) target. For the vector mediator (bottom panel) three different exposure have been considered: 1 kg-yr, 10 kg-yr, and 100 kg-yr. The current excluded regions  from XENONnT (yellow region) and Gemma/Borexino (orange region) are also shown.  We have assumed a constant background of 0.1 DRU and the detector threshold is 5 eV. }
\end{figure}

\begin{figure}[tbp]
\centering 
\includegraphics[width=.45\textwidth]{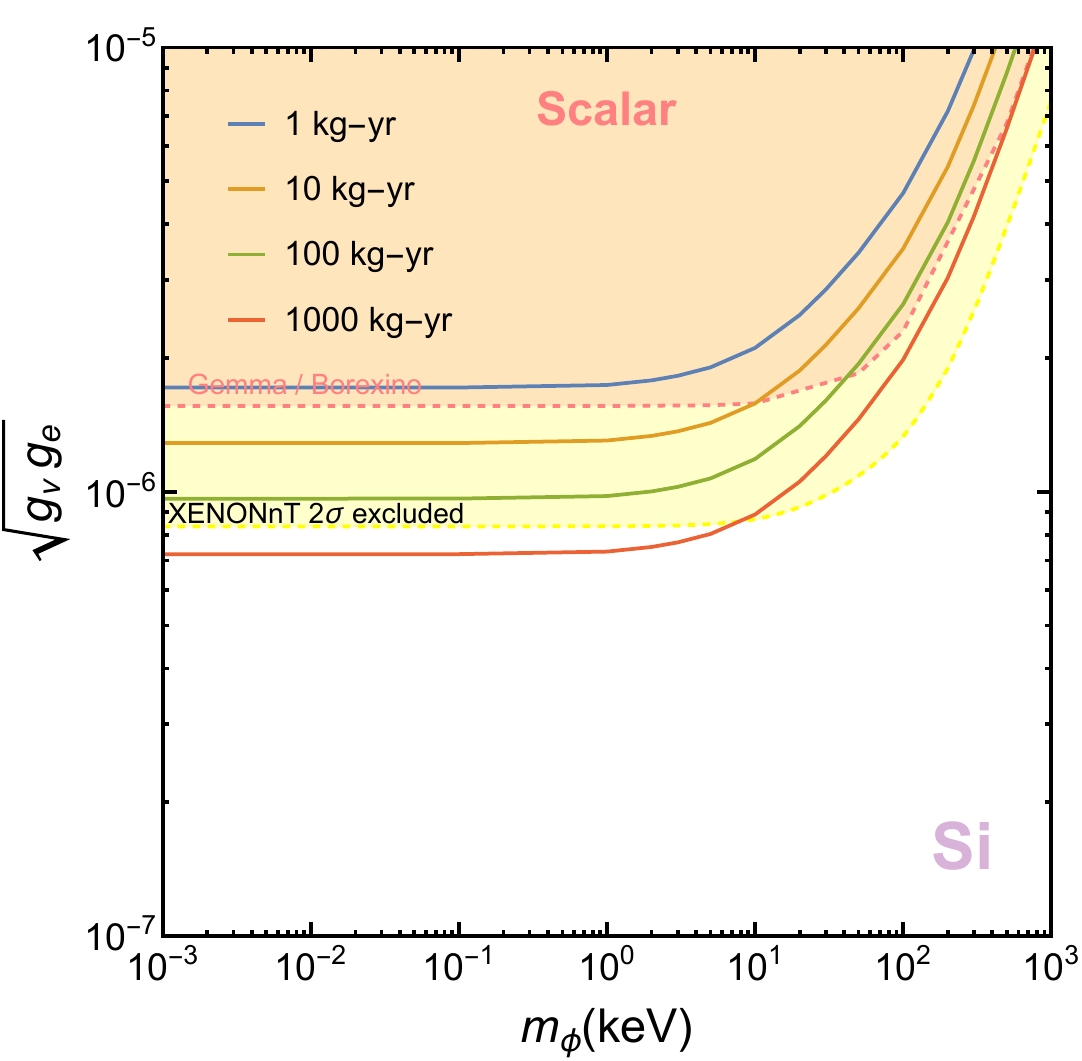}
\hfill
\includegraphics[width=.45\textwidth]{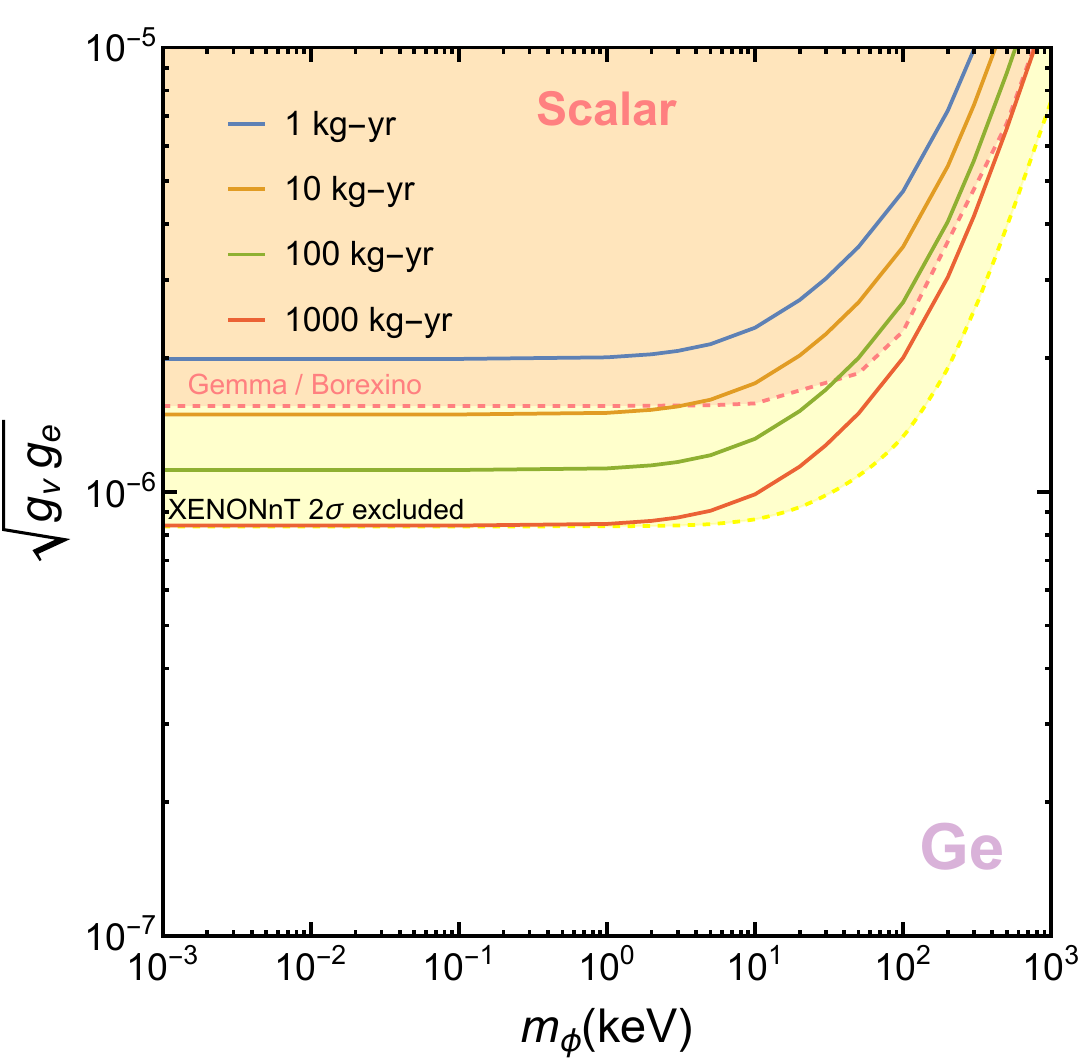}
\hfill
\includegraphics[width=.45\textwidth]{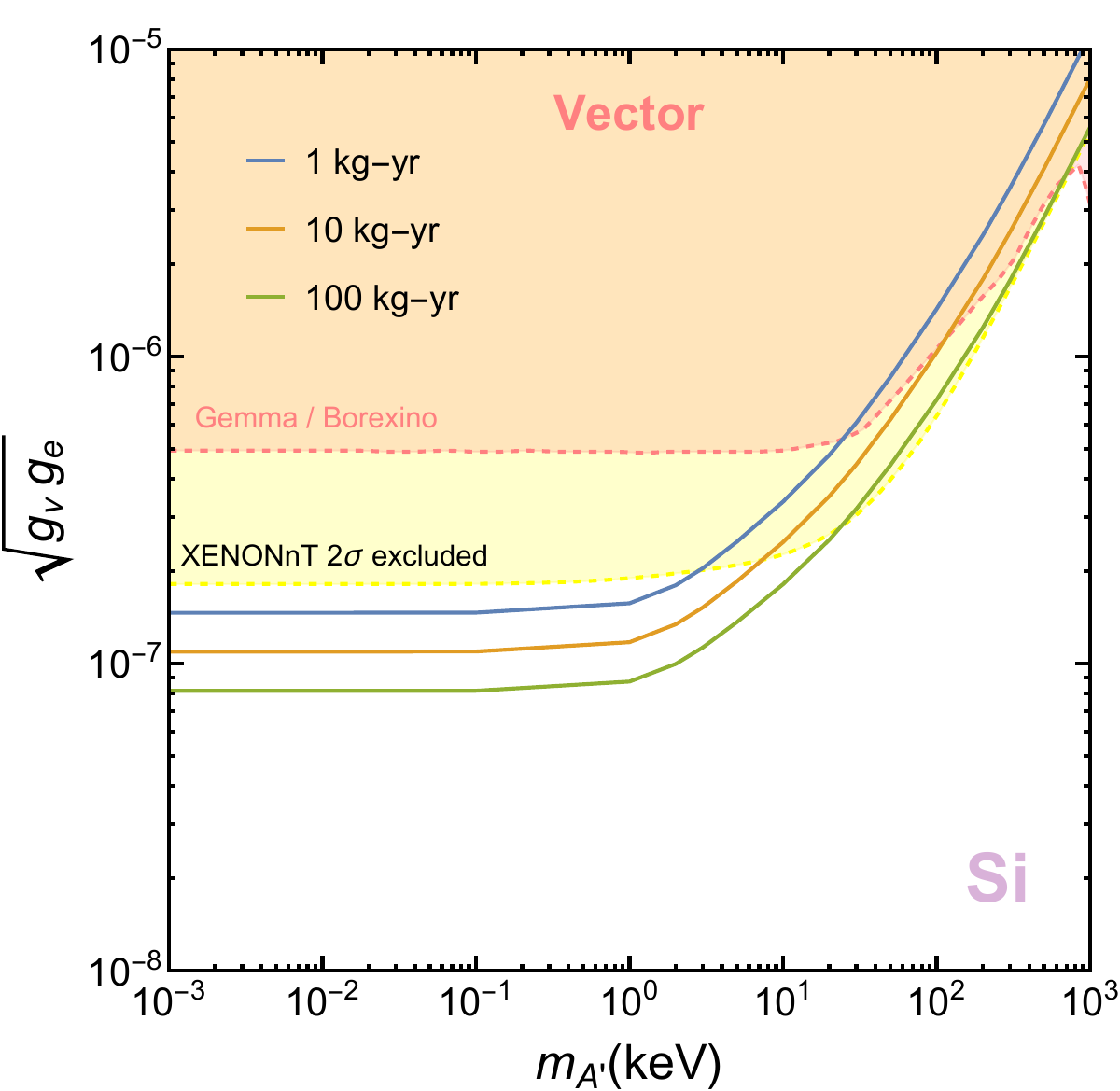}
\hfill
\includegraphics[width=.45\textwidth]{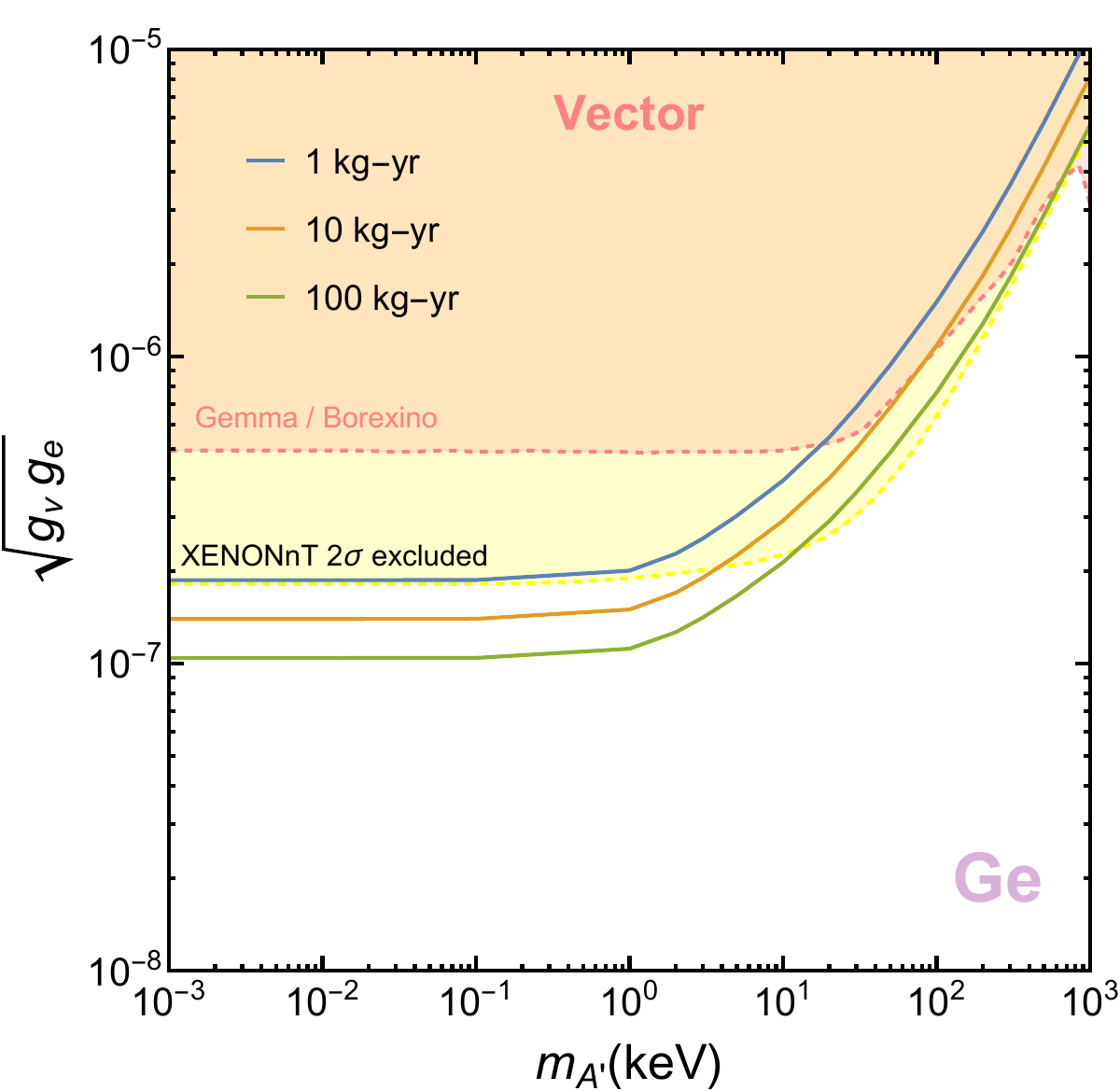}
\captionsetup{justification   = RaggedRight,
             labelfont = bf}
\caption{\label{fig:sensitivity1dru}Projected bounds on the coupling $\sqrt{g_\nu g_e}$ at $90 \%$ confidence level as a function of the mediator masses assuming a constant background of 1 DRU and the detector threshold is 5 eV.. The bounds are given for four different exposure values: 1 kg-yr, 10 kg-yr, 100 kg-yr, and 1000 kg-yr for scalar mediators (top panel) for silicon (left) and germanium (right) target. For the vector mediator (bottom panel) three different exposure have been considered: 1 kg-yr, 10 kg-yr, and 100 kg-yr. The current excluded regions  from XENONnT (yellow region) and Gemma/Borexino (orange region) are also shown. }
\end{figure}

In figure~\ref{fig:sensitivitypoint1dru} we show the projected sensitivity at 90\% confidence level for both scalar and vector mediator models considering a constant background of 0.1 DRU over the SM $\nu-e$ background. Along with this we also show the excluded region of parameter space from GEMMA, Borexino, and XENONnT experiments. The constraints from CONUS and CONNIE are subdominant. A similar plot for a constant background of 1.0 DRU is shown in figure~\ref{fig:sensitivity1dru}. As expected we see an enhanced sensitivity to light mediators using a detector with low threshold. The nature of the exclusion curve can be described as follows;~\begin{itemize}
    \item For heavier mediator masses the coupling sensitivity decreases since for $q<<M_{\phi/A'}$ the new-physics interaction becomes the Fermi like contact interaction. This gives a line with a slope of 2 in the log-log space of the coupling vs mediator mass plot.
    \item For light mediators, the mediator mass may be neglected in the propagator when it becomes small with respect to the momentum transfer $q>>M_{\phi/A'}$ which corresponds to the sensitivity becoming flat for very light mediators.
\end{itemize}

We notice that the larger flux of the reactor experiment allows for a deeper reach into the parameter space when compared to solar neutrino experiments. Even a 1 year run of a 1 kg Si or Ge detector can  probe larger area of the parameter space compared to the Gemma, Borexino, and XENONnT experiments. This shows that a reactor based experiment using low threshold detectors capable of differentiating ER-NR signals like those intended for MINER can provide prominent bounds on large regions of parameter space. Note that one can reach an even greater sensitivity at a GW class reactor experiment but there one is limited by the distance the detector can be placed from the reactor core which is typically $\sim$10 m. Overall, the flux is still enhanced by an order of magnitude at GW reactor experiments.


\begin{figure}[tbp]
\centering 
\includegraphics[width=.45\textwidth]{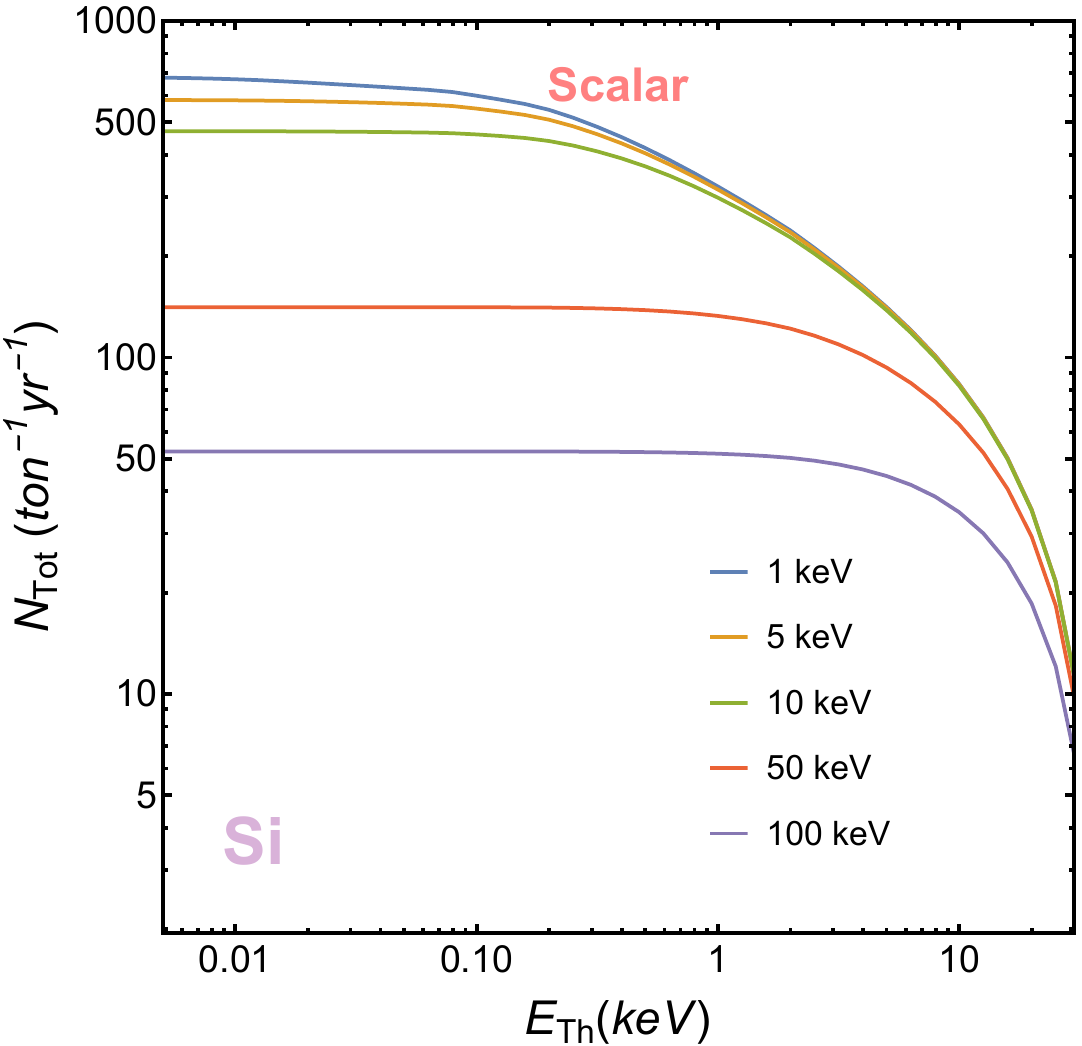}
\hfill
\includegraphics[width=.45\textwidth]{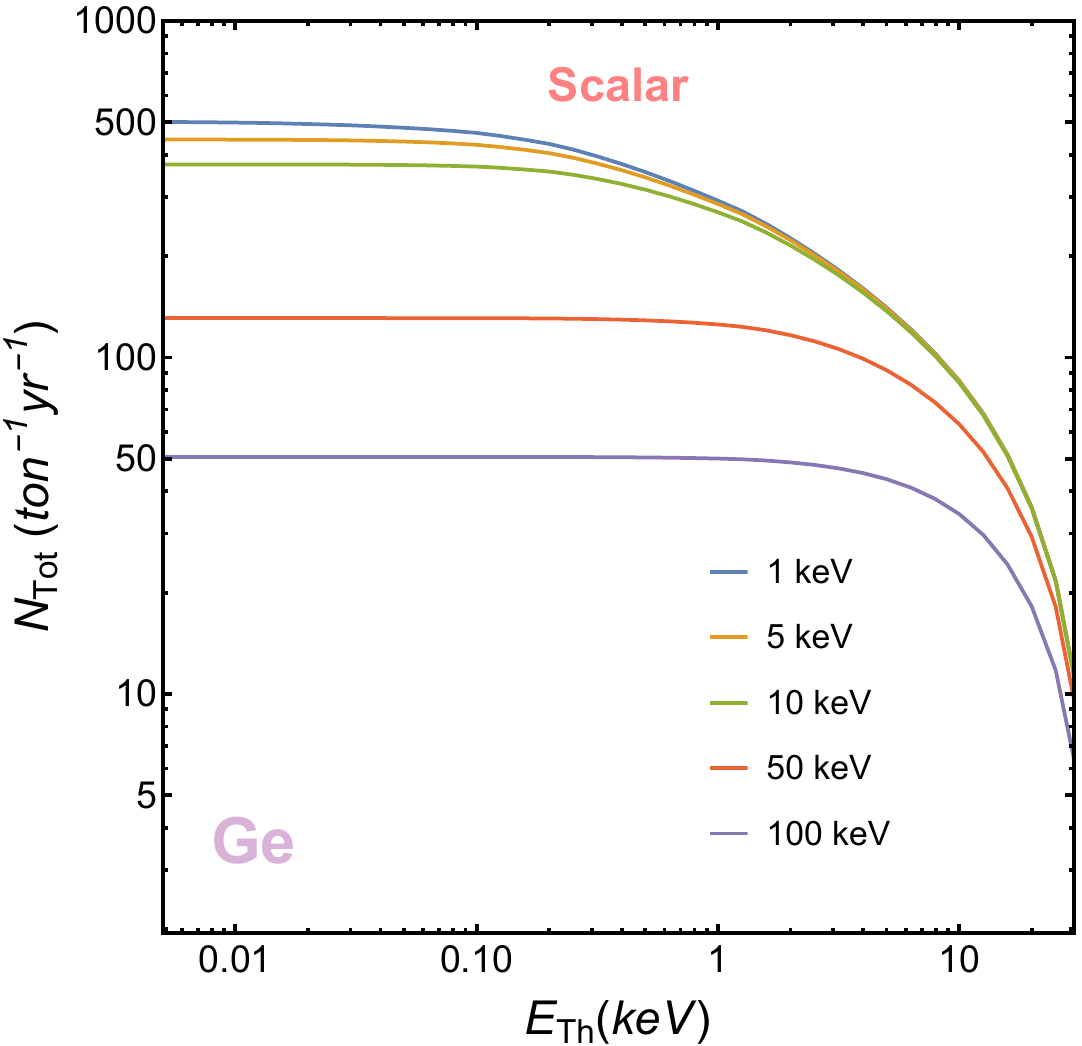}
\hfill
\includegraphics[width=.45\textwidth]{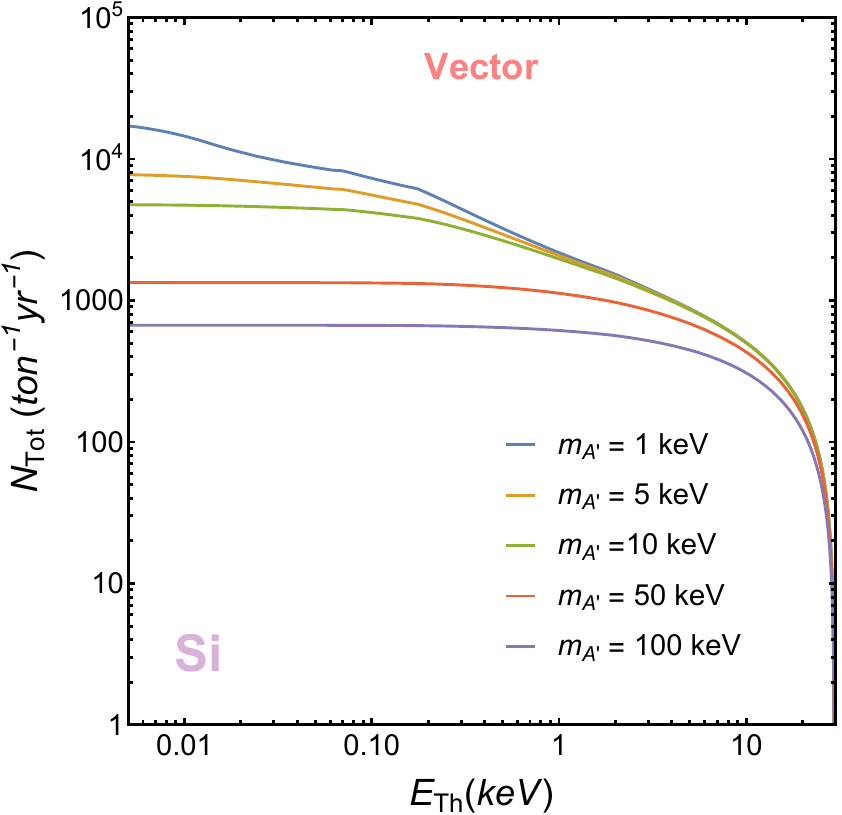}
\hfill
\includegraphics[width=.45\textwidth]{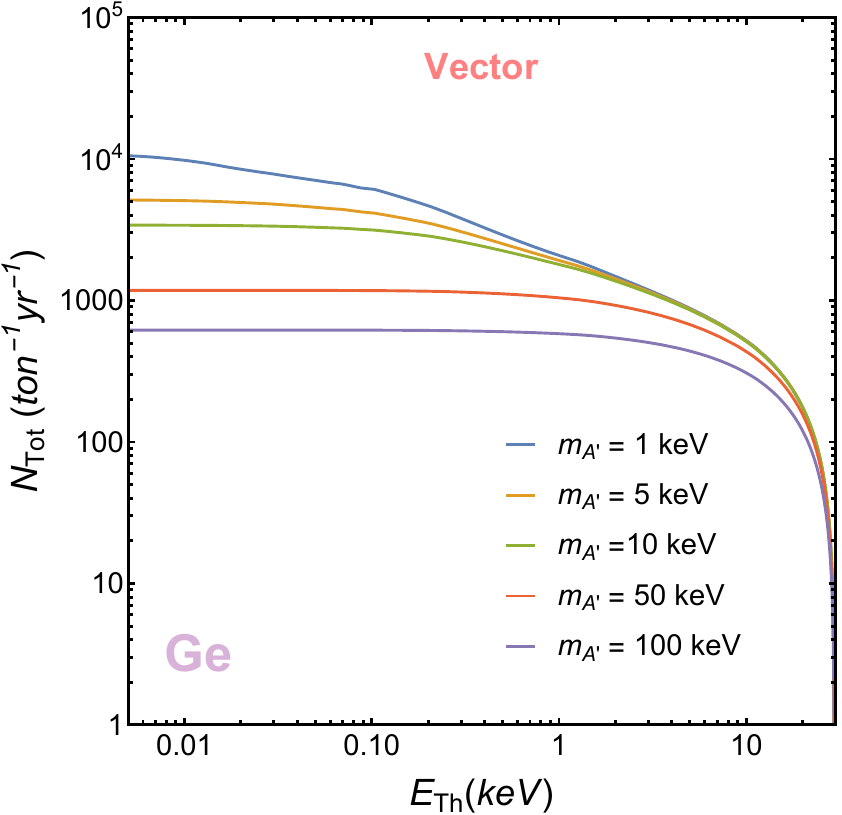}
\captionsetup{justification   = RaggedRight,
             labelfont = bf}
\caption{\label{fig:NTot} Integrated yield of electron recoil events (per ton per year) captured by silicon and germanium detectors as a function of detector threshold $E_{Th}$. The coupling values are taken as $g_{\nu}=g_{e}=5 \times 10^{-7}$ for scalar mediator (Top Panel), and
$g_{\nu}=g_{e}=1.5 \times 10^{-7}$ for vector mediator (Bottom Panel).}
\end{figure}

In figure~\ref{fig:NTot} we show the projected total number of events per ton per year as a function of experimental recoil threshold assuming the vector and scalar couplings to be $g_{\nu}=g_{e}=1.5\times10^{-7}$ and $5\times10^{-7}$, respectively. The electron recoil events for both Si and Ge detector targets for various mediator masses are plotted. As expected, in all the cases shown, lowering the threshold increases the reach of the experiment for light mediator masses. We see that for scalar case a $\sim$100 eV threshold  is good enough to capture most of the events, though an even lower threshold can help provide a better resolution of the light mediator mass. On the other hand, for the vector case a $\sim$1 eV threshold detector can capture an order of magnitude more events compared to  a $\sim$1 keV threshold detector for $\mathcal{O}(1)$~keV mediator mass . This implies that the valence to conduction transition rate calculation used in this work is particularly significant for these very light vector mediators.

\section{Conclusions} \label{sec:conclusion}

The non-standard neutrino-electron interactions via light and weakly coupled scalar or vector mediators occur in many extensions of the  SM such as $U(1)_{L_i-L_j}$, $U(1)_{B-L}$, and $U(1)_{T3R}$, etc. But not many such scalar mediator models exist. Here we proposed one light scalar model by extending only the scalar sector of the SM using doublet, triplet, and singlet scalar fields. In this model, we have non-standard interactions associated with different flavors of neutrinos for both heavy and light mediators. Results of many previous experiments utilizing solar and reactor flux such as XENONnT, Borexino, GEMMA, etc  provide constraints on the parameter space of such light mediator models  associated with non-standard $\nu_e-e$ scattering.

 Since MW class reactor experiments have an order of magnitude more neutrino flux (at a mean distance of 1m from the core) compared to solar experiments, reactor experiments are expected to provide greater sensitivity. Ongoing and future reactor experiments such as MINER, CONUS, CONNIE, vIOLETTA, etc. can provide a particularly important probe of such light mediator models  if one is to leverage ultra low threshold detectors such as Si and Ge. We discussed the prospects of probing non-standard neutrino interactions with low energy neutrino-electron elastic scattering in reactor based experiments. Since for these neutrino-electron interactions via light mediators, the typical momentum transfer is low, the energy deposited in the detector is in the sub-keV range. At these sub keV energies, one cannot reliably use the free electron approximation (FEA) and must include the atomic and crystal effects in the calculation. Taking this into account via atomic form factor from \texttt{DarkARC} and valence to conduction crystal form factor from \texttt{EXCEED-DM} we evaluated the sensitivity of a typical MW-class reactor experiments to weakly coupled NSIs in light mediator models.
We showed that using ultra low threshold detectors such as Si/Ge the reactor experiments can outdo the current constraints on light scalar and vector mediators couplings to neutrinos and electrons.  We also find that a low threshold detector with very low background capable of differentiating between ER and NR is very crucial for such deeper reach of the parameter space. A  germanium (silicon) detector with $10$~kg-yr exposure and 1 MW reactor anti-neutrino flux would be able to probe a scalar and vector mediator with masses below 1 keV for coupling products $\sqrt{g_\nu g_e}$ $\sim$ $1 \times 10^{-6}~(9.5 \times 10^{-7}) ~{\rm and}~ 1\times 10^{-7} ~(8\times 10^{-8})$, respectively.

\acknowledgments

We are grateful to Timon Emken, C.-P. Liu, Rupak Mahapatra, Tanner Trickle, P. S. Bhupal Dev and Miguel Escudero for useful discussions. The work of BD, SG, AT and AV is supported in part by DOE grant DE-SC0010813. The work of SG is also supported in part by National Research Foundation of Korea (NRF) Grant No. NRF-2019R1A2C3005009 (SG). T.L. is supported in part by the National Key Research and Development Program of China Grant No. 2020YFC2201504, by the Projects No. 11875062, No. 11947302,  No. 12047503, and No. 12275333 supported by the National Natural Science Foundation of China, as well as by the Key Research Program of the Chinese Academy of Sciences, Grant NO. XDPB15. We acknowledge that portions of this research were conducted with the advanced computing resources provided by Texas A\&M High Performance Research Computing.

\paragraph{Note added.} We provide the files containing the codes for the numerical calculations here: \href{https://github.com/averma609/NSI-calc}{https://github.com/averma609/NSI-calc}

\bibliographystyle{JHEP.bst}
\bibliography{NSI.bib}

\end{document}